\documentclass[
    authoryear,  %
    5p,
    nonatbib, %
]{elsarticle}

\usepackage[utf8]{inputenc}
\usepackage[english]{babel}
\usepackage[todonotes={obeyFinal}]{changes}
    \definechangesauthor[color=teal]{BvH}
    \definechangesauthor[color=cyan]{JJ}
    \definechangesauthor[color=red]{MAP}
    
\usepackage[
    pdftitle={Dendritic structure enables powerful plasticity},
    pdfauthor={NeuroTMA},
    colorlinks=true,
    ]{hyperref}
\usepackage[
    backend=biber,
    style=authoryear,
    sorting=nyt,
    useprefix=true,  %
    maxbibnames=10,
    minbibnames=10,
    maxcitenames=2,
    mincitenames=1,
    dashed=false,
    giveninits=true,  %
    uniquelist=minyear,  %
    uniquename=minyearinit,  %
]{biblatex}
    \DeclareNameAlias{sortname}{family-given}
        \renewbibmacro*{doi+eprint+url}{%
            \iftoggle{bbx:url}
            {\iffieldundef{doi}{\usebibmacro{url+urldate}}{}}
            {}%
            \newunit\newblock%
            \iftoggle{bbx:eprint}
            {\usebibmacro{eprint}}
            {}%
            \newunit\newblock%
            \iftoggle{bbx:doi}
            {\printfield{doi}}
            {}
        }
        \newcommand*{\clearisXn}{%
          \iffieldundef{doi}%
            {\iffieldundef{url}%
                {}{\clearfield{issn}\clearfield{isbn}}%
            }%
            {\clearfield{issn}\clearfield{isbn}}}
        \AtEveryBibitem{\clearisXn}
        \AtEveryCitekey{\clearisXn}
    \DeclareSourcemap{
      \maps[datatype=bibtex, overwrite]{
          \map{
              \step[fieldset=editor, null]
          }
      }
    }

\usepackage{amsmath}
\usepackage{amssymb}
\usepackage{bm}
\usepackage[labelfont=bf]{caption}
\usepackage[
    capitalize,
    nameinlink,
    poorman,
]{cleveref}
\usepackage{csquotes}
\usepackage{enumitem}
\usepackage{etoolbox}  %
\usepackage{geometry}
\usepackage[
    docdef=atom,
    nomain,
    acronym,
    automake,
    nonumberlist,  %
    nogroupskip,  %
]{glossaries-extra}
\makeglossaries{}
    \setabbreviationstyle[acronym]{long-short-user}
    \glssetcategoryattribute{acronym}{nohyperfirst}{true}
    \newignoredglossary*{ignored} %
\newacronym{adex}{AdEx}{adaptive exponential leaky integrate-and-fire}
\newacronym{ann}{ANN}{artificial neural network}
\newacronym{am}{AM}{adjoint method}
\newacronym{bccn}{BurstCCN}{bursting cortico-cortical networks}
\newglossaryentry{burstprop}{
    type={ignored},
    name={Burstprop},
}
\newacronym{bp}{BP}{error backpropagation}
\newacronym{bptt}{BPTT}{backpropagation through time}
\newacronym{btsp}{BTSP}{behavioral timescale synaptic plasticity}
\newglossaryentry{ca3}{
    name={CA3},
    description={cornu ammonis area 3},
}
\newacronym{cmc}{CMC}{cortical microcircuit}
\newglossaryentry{cnn}{
    name={CNN},
    description={convolutional neural network},
}
\newacronym{dcmc}{dendritic CMC}{dendritic cortical microcircuit}
\newacronym{dbp}{dendritic BP}{dendritic backpropagation}
\newacronym{dfc}{DFC}{deep feedback control}
\newacronym{dg}{DG}{dentate gyrus}
\newacronym{dhpc}{dendritic hPC}{dendritic hierarchical predictive coding}
\newacronym{dtp}{dendritic TP}{dendritic target propagation}
\newacronym{ec3}{EC3}{entorhinal cortex layer 3}
\newacronym{et}{ET}{eligibility trace}
\newacronym[description={\glsfmtlong{elise} (model)}]{elise}{ELiSE}{efficient learning of sequences}
\newacronym{enmc}{ENMC}{error neuron microcircuit}
\newacronym{fa}{FA}{feedback alignment}
\newglossaryentry{follow}{
    name={FOLLOW},
    description={feedback-based online local learning of weights},
}
\newacronym{le}{LE}{latent equilibrium}
\newacronym{gle}{GLE}{generalized latent equilibrium}
\newacronym{lif}{LIF}{leaky integrate-and-fire}
\newacronym{lman}{LMAN}{lateral magnocellular nucleus of the ant.\ nidopallium}
\newacronym{ltd}{LTD}{long-term depression}
\newacronym{ltp}{LTP}{long-term potentiation}
\newacronym{mc}{MC}{microcircuit}
\newacronym{ml}{ML}{machine learning}
\newacronym{mse}{MSE}{mean squared error}
\newacronym{nlif}{nLIF}{non-leaky integrate-and-fire}
\newglossaryentry{nmda}{
    type={ignored},
    name={NMDA},
    description={\textit{N}-methyl-D-aspartate},
}
\newacronym{ode}{ODE}{ordinary differential equation}
\newacronym{ostl}{OSTL}{online spatio-temporal learning}
\newacronym{pal}{PAL}{phaseless alignment learning}
\newacronym{pc}{PC}{predictive coding}
\newacronym[description={\glsfmtlong{pid} (controller)}]{pid}{PID}{proportional-integral-derivative}
\newacronym[type=ignored]{psp}{PSP}{postsynaptic potential}
\newacronym[description={\glsfmtlong{pv} (interneuron)}]{pv}{PV}{parvalbumin-positive}
\newglossaryentry{ra}{
    name={RA},
    description={robust nucleus of the arcopallium},
}
\newacronym{rflo}{RFLO}{random feedback online learning}
\newacronym{rnn}{RNN}{recurrent neural network}
\newacronym{rtrl}{RTRL}{real-time recurrent learning}
\newacronym{sal}{SAL}{spike-based alignment learning}
\newacronym[type=ignored]{sbi}{SBI}{simulation-based inference}
\newacronym{snn}{SNN}{spiking neural network}
\newacronym{srm}{SRM}{spike response model}
\newacronym[description={\glsfmtlong{sst} (interneuron)}]{sst}{SST}{somatostatin-positive}
\newacronym{stdp}{STDP}{spike-timing-dependent plasticity}
\newacronym{std}{STD}{short-term depression}
\newacronym{stf}{STF}{short-term facilitation}
\newacronym{tp}{TP}{target propagation}
\newacronym{ttfs}{TTFS}{time-to-first-spike}
\newacronym{upe}{UPE}{uncertainty-modulated prediction error}
\newacronym[description={\glsfmtlong{us} (learning rule)}]{us}{US}{Urbanczik \& Senn}
\newacronym[description={\glsfmtlong{vip} (interneuron)}]{vip}{VIP}{vasoactive intestinal peptide}
\newacronym{vle}{VLE}{variational latent equilibrium}

\usepackage{graphicx} %
\usepackage{multicol}
\usepackage[
    group-minimum-digits=4,  %
]{siunitx}
\usepackage{tabularx}
\usepackage[most]{tcolorbox}
    \newcounter{counterOfBoxes}
    \crefname{counterOfBoxes}{Box}{Boxes}
    \newtcolorbox[use counter=counterOfBoxes]{boxWithCounter}[2]{
        enhanced,
        float*=t,
        width=\textwidth,
        label = #1,
        title ={Box~\thecounterOfBoxes:~#2},
    }
\usepackage[titles]{tocloft}
\usepackage[obeyFinal]{todonotes}

    \makeatletter
    \def\@footnotecolor{red}
    \define@key{Hyp}{footnotecolor}{%
     \HyColor@HyperrefColor{#1}\@footnotecolor%
    }
    \def\@footnotemark{%
        \leavevmode
        \ifhmode\edef\@x@sf{\the\spacefactor}\nobreak\fi
        \stepcounter{Hfootnote}%
        \global\let\Hy@saved@currentHref\@currentHref
        \hyper@makecurrent{Hfootnote}%
        \global\let\Hy@footnote@currentHref\@currentHref
        \global\let\@currentHref\Hy@saved@currentHref
        \hyper@linkstart{footnote}{\Hy@footnote@currentHref}%
        \@makefnmark
        \hyper@linkend
        \ifhmode\spacefactor\@x@sf\fi
        \relax
      }%
    \makeatother

    \makeatletter\renewcommand{\todo}[2][]{\@bsphack \@esphack } \makeatother
\usepackage{xspace}

\newcommand{\Cm}{C\_m}

\newcommand{\El}{E_\leak}
\newcommand{\gsd}{g_\sd}
\newcommand{\gl}{g_\leak}

\newcommand{\leak}{\mathrm{l}}

\newcommand{\pre}{\mathrm{pre}}
\newcommand{\rev}{\mathrm{rev}}
\newcommand{\rpre}{r_\pre}
\newcommand{\rprebar}{\bar r_\pre}
\newcommand{\sd}{\mathrm{sd}}

\newcommand{\taum}{\tau\_m}
\newcommand{\taur}{\tau\_r}

\newcommand{\ubrever}{\breve{u}^\mathrm{r}}
\newcommand{\WT}{W^\mathrm{T}}
\newcommand*\subtxt[1]{_{\textnormal{#1}}}
\DeclareRobustCommand\_{\ifmmode\expandafter\subtxt\else\textunderscore\fi}

\title{Dendritic structure enables powerful plasticity}
\author[1]{Ben von Hünerbein\corref{cor1}}
\author[1]{Federico Benitez\corref{cor1}}
\author[2]{Kevin Max}
\author[1]{Julian Göltz}
\author[1]{Paul Haider}
\author[1]{Simon Brandt}
\author[3]{Arno Granier}
\author[1]{Timo Gierlich}
\author[1]{Jakob Jordan}
\author[4]{Katharina A. Wilmes}
\author[1]{Jean-Pascal Pfister}
\author[1]{Walter Senn}
\author[1]{Mihai A. Petrovici}

\cortext[cor1]{Shared first authorship.}

\affiliation[1]{organization={Department of Physiology, University of Bern},
city={Bern},
country={Switzerland}}
\affiliation[2]{organization={Neural Computation Unit, Okinawa Institute of Science and Technology},
city={Onna},
country={Japan}}
\affiliation[3]{organization={Département d’Études Cognitives,
École Normale Supérieure - PSL},
city={Paris},
country={France}}
\affiliation[4]{organization={Institute of Neuroinformatics, University of Zürich and ETH Zürich},
city={Zürich},
country={Switzerland}}

\begin{document}

\newgeometry{
    left=1.6cm,
    right=1.1cm,
    top=2.5cm,
    bottom=2.5cm,
}

\hypersetup{
    linkcolor=blue,
    citecolor=teal,
    filecolor=black,
    urlcolor=cyan,
    footnotecolor=orange,
}

\begin{abstract}
Over the past decades, it has become increasingly clear that the complex morphology of cortical neurons is more than just a quirk of evolution, and that dendritic compartments serve as computational elements in their own right, rather than just providing connections between nerve cell bodies.
While most computational studies discuss the enhanced representational capabilities of multi-compartment models as compared to point neurons, we focus here on the implications of neuronal morphology for synaptic plasticity.
We argue that the ability of single neurons to simultaneously encode multiple pieces of information gives synapses local access to more than just the classical Hebbian pre- and postsynaptic terms, and with much greater specificity and reaction speed than permitted by other globally modulated factors.
Based on a comparative review of recent dendritic learning models, we show how such neuronal compartmentalization can provide synapses with the means for calculating various forms of error signals, which in turn give rise to powerful real-time and fully local instantiations of deep learning through gradient descent.
Implemented within cortical microcircuits capable of propagating and manipulating these errors, compartmentalized neurons thus ultimately enable the learning of far more complex tasks than are achievable by globally modulated Hebbian plasticity alone.
\end{abstract}

\begin{keyword}
dendritic compartments \sep error-correcting plasticity \sep biological deep learning \sep cortical microcircuits
\end{keyword}

\maketitle

\setcounter{tocdepth}{2} %
\makeatletter \renewcommand{\@dotsep}{10000} \makeatother  %
\renewcommand{\cftdotsep}{10000}
{ %
  \hypersetup{linkcolor=black}
  \setlength{\cftbeforesecskip}{4pt} %
  \tableofcontents
}

\section{Introduction}

\begin{sloppypar}
It is a common tenet in biology that evolutionary pressure often leads to highly optimized organisms.
However, features may appear as mere byproducts \textendash{} spandrels~\parencite{gould1979spandrels}
\textendash{} of other adaptive characteristics or may vestigially persist over evo\-lu\-tion\-ary time\-scales even when obsolete or clearly suboptimal, such as the infamous laryngeal nerve of the giraffe.
Thus, the mere presence of individual features cannot in itself constitute a proof of their optimality, and even less so provide an account of their specific functional role.
\end{sloppypar}

The complex morphology of neurons in the brain represents a prominent feature which has raised such questions about functionality, garnering increasing attention over the past few decades.
While dendritic trees are obviously required to connect single cells to a large number of presynaptic partners, and fulfill a role as cables for (electric) signal transport \parencite{haeusser2003dendrites, larkum2022are}, the interesting question relates to their \emph{computational} role in brain function.

By now, it has become irrefutable that dendrites are not merely passive components, but play an active role in neuronal computation.
In light of the ample evidence for various mechanisms of nonlinear dendritic processing, our picture of single neurons has undergone a radical shift, from simple point models to the equivalent of entire multilayer, or even deep networks \parencite{haeusser2003dendrites, poirazi2003pyramidal, gidon2020dendritic, beniaguev2021single}.
 However, while this explains how dendrites can be computationally \emph{useful}, it does not account for why they might be computationally \emph{necessary}.
Indeed, the equivalence to deep \glspl{ann} could be taken as a demonstration of the opposite, as these works do not show how networks of structured neurons can go in any way beyond what \glspl{ann} are already capable of doing with point neurons alone.
The question of the necessity of dendrites has important implications not only for computational models of the brain, but also for artificial, brain-inspired computing, as it shapes our understanding of the biological primitives of computation.

Here, we provide a different perspective on the role of dendrites, which ultimately highlights their role as necessary, rather than merely useful components for complex computation.
Specifically, we suggest that their ability to locally and simultaneously track multiple instructive signals enables powerful local synaptic plasticity.
To set up this perspective, let us return to \glspl{ann} and the workhorse of deep learning: \glsfmtlong{bp} \parencite[][]{linnainmaa1970representation,werbos1982applications,rumelhart1986learning}.
While undoubtedly effective, \gls{bp} is manifestly nonlocal in space and time, as rightfully remarked by early critics of \gls{bp} as a model for learning in the brain \parencite{crick1989recent}.
Nonlocality in space refers to the transport of distant errors to individual synapses, which requires feedback information to pass through feedforward synapses \textendash{} otherwise known as the weight transport problem.
Nonlocality in time refers to error signals only becoming available after activity has fully propagated throughout a network, at a time when they might no longer match the currently present activity.
These issues are only exacerbated in \glsfmtlong{bptt} \parencite[\glsfmtshort{bptt};][]{pineda1987generalization,werbos1990backpropagation}, where activity needs to be played back through the same network, along with the calculation of errors, but in reverse time.
Thus, in absence of additional mechanisms, the functional proficiency attainable by networks of point neurons under \gls{bp} remains inaccessible to biological neuronal networks.

For contrast, we can consider popular learning rules in classical computational neuroscience.
The simplest such rule is due to Hebb and adapts synapses solely based on pre- and post-synaptic activity~\parencite{hebb1949organization}.
Prominent early work showed that networks instilled with Hebbian plasticity can learn to perform relatively complex computational tasks.
Some well-known examples include associative memory learning~\parencite{hopfield1982neural}, self-organizing behavior~\parencite{kohonen1998selforganizing}, principal component analysis upon inclusion of a simple homeostatic term~\parencite{oja1982simplified}, and
sampling-based Bayesian inference~\parencite{hinton1986learning,hinton2002training}.
However, these capabilities remain somewhat limited, especially in comparison with modern \glspl{ann} powered by deep learning.

Significant additional power can be achieved by extending classical Hebbian plasticity with a `third factor', which modulates learning based on a globally available signal quantifying novelty, reward, or errors \parencite[reviewed extensively in][]{fremaux2016neuromodulated, kusmierz2017learning, gerstner2018eligibility}.
This has been shown to enable learning of patterns \parencite{brea2013matching, jimenezrezende2014stochastic}, aggregate label learning of speech \parencite{guetig2016spiking}, and blind source separation of images and movies \parencite{isomura2016local}.
However, such third factors usually represent the effect of neuromodulators whose dynamics are governed by slow diffusion processes, which tend to act slowly and globally, with very limited spatial and temporal specificity, strongly limiting their capabilities~\parencite{gerstner2018eligibility}.

At this point, the next natural step appears to be the inclusion of additional learning rule components that have a greater degree of specificity.
If these components are to evolve on the same time scales as neuronal signals, while also relating specifically to individual neurons, then the most natural physical locus of such components would be within the neurons themselves.
Dendritic trees provide the required subdivision of neurons into compartments for separately storing these components.

In this perspective article, we posit the view that the addition of dendritic compartments represents a \emph{necessary} prerequisite for the instantiation of powerful, fully local, and phase-free learning.
Thus, the search for a unique function of dendritic trees meets the search for a biologically plausible substrate for deep learning, and two major outstanding challenges directly answer each other's call.

Along with fostering a better mechanistic understanding of computation and learning in biological brains, multi-compartment models also provide interesting components for the development of artificial brains.
While superficial observation might
suggest that effective cross-pollination between artificial and biological neural networks culminates
with receptive fields~\parencite{hubel1959receptive,hubel1962receptive} and their applications in filtering~\parencite{fukushima1980neocognitron} and \glspl{cnn}~\parencite{lecun1989backpropagation,lecun1998gradientbased,krizhevsky2012imagenet}, \glspl{ann} still have quite the gap to bridge towards their biological paragon, regarding both functional capabilities and the associated resource consumption.
This suggests that there is still enormous potential for transfer of knowledge, especially with respect to the neuro-synaptic dynamics underlying computation.
In particular, the remarkable levels of performance achieved by \glspl{ann} are only permitted by energy consumptions that exceed those of our brains by several orders of magnitude.
Much of this comes down to information transport between different hardware components made necessary by nonlocal operations.
The manifest locality of synaptic learning rules enabled by dendritic structure
is set to close this gap.
Compartmental models and associated plasticity rules are thus especially promising for neuromorphic system design.

What follows is a systematic overview of neuronal plasticity models that use dendritic compartments to enable powerful synaptic learning rules.
Our specific focus is on error-correcting plasticity and how it can be facilitated by dendrites.
In order to organize the discussion, we separate these models into two classes, emphasizing error representation.
Specifically, we differentiate between
models based on whether errors are represented \emph{explicitly} in a dendritic compartment, as opposed to models where errors are only \emph{implicitly} driving neuronal dynamics.
The difference is easy to visualize from an experimental point of view:
Models with explicit error representation are those in which the errors are represented in specific neuronal compartments, which could in principle be measured directly by a suitable probe.
Conversely, in models with implicit error representation,
errors still drive plasticity, but they are computed in the synapse from other postsynaptic signals.
Thus, no experimental setup could probe a specific compartment to directly measure the value of the error signal.
In addition to the explicit and implicit representation of \emph{errors}, we also discuss a third class of models that take data \emph{uncertainty} into account.
Because sensory data is inherently variable, incorporating this information can be beneficial for learning, and compartmentalized neurons enable a corresponding modulation of synaptic plasticity rules.

In this work we do not explicitly distinguish between rate- and spike-based
models, but rather focus on well-defined learning schemes with concrete experimental correlates and measurable metrics of functionality for a wide array of different models.
We refer the reader to \cref{box:modelBackground} for an introduction to the basic elements and notations common to all models.

\begin{boxWithCounter}{box:modelBackground}{Modeling background}
    Variables and equations can either describe single neurons or population of neurons as $u\to\bm{u}$.
    Correspondingly, connections can be read as interactions between single neurons or between populations as $W\to\bm{W}$.
    Signals $x \in \{r,e\}$coming from lower areas in the cortical hierarchy are denoted by ${x}^\downarrow$ and signals coming from higher areas by ${x}^\uparrow$.

    \begin{multicols}{2}
        \vspace{1em}
        \begin{center}
            \textbf{Definitions}
        \end{center}

        \vspace{1em}
        \begin{tabularx}{0.95\linewidth} {
                c
                >{\raggedright\arraybackslash}X
             }
             $u$
             &
             \textbf{membrane voltage} - {the state variable of a neuron that changes with synaptic input}
             \\
             $v$
             &\textbf{dendritic voltage} - {the state variable of a neuron's dendritic membrane}
             \\
             $\taum$
             &
             \textbf{membrane time constant} - {defines how fast a neuron reacts to input}
             \\
             $r$
             &
             \textbf{neuronal output rate} - {usually a function of ${u}$ with activation $\varphi$ such that $r(u)=\varphi(u)$, while some models use prospective outputs with $r(u, \dot{u})=\varphi(\breve u)$ (see \Cref{box:preprospective})}
             \\
             $e$
             &
             \textbf{error} - {many models explicitly encode errors that can be computed by dendrites or specific error neurons}
             \\
             $W$
             &
             \textbf{synaptic weights} - {connection strengths between pairs of neurons}
             \\
             $\eta$
             & \textbf{learning rate} - {controls the amplitude of weight updates and thereby the speed of plasticity; often omitted for brevity}
        \end{tabularx}

        \vspace{1em}
        \begin{center}
            \textbf{Important concepts}
        \end{center}

        \vspace{-0.0em}
        \begin{description}[style=nextline]
            \item[Multi-compartment leaky integration]
                \[
                  \begin{aligned}[t]
                    \Cm\dot{u} &= \gl(\El - u) + Wr\,\gamma + G(v - u) \\[4pt]
                    \Cm\dot{v} &= \gl(\El - v) + Wr\,\gamma
                  \end{aligned}
                \]
        \\{
        with membrane capacitance $\Cm$, leak conductance $\gl$, and leak reversal potential $\El$.
        \columnbreak
        Some models ignore the dendritic leak.
        ${Wr\,\gamma}$ represents synaptic currents with weight matrix ${W}$ and presynaptic rate vector ${r}$,
        and ${G(v - u)}$ represents coupling to other compartments
        with coupling conductance matrix ${G}$.
        The factor $\gamma$ accounts for the type of synaptic interaction, with ${\gamma=1}$ for current-based and
        ${\gamma=E^\rev-u}$ for conductance-based synapses.
        }
            \item[Delta rule:\quad $\Delta W=\eta (r^* - r) \rpre$]
            synaptic weight updates depend on the presynaptic rate $\rpre$ and the difference (delta) between the current neuronal output $r$ and a target activity $r^*$.
            Most rules described here are a variant of the delta rule.

            \item[Error propagation:\quad $e=r'\cdot \WT e^\uparrow$]
            This variant of error propagation is used in modern AI algorithms and approximated by some models in this manuscript. Errors from upper layers $e^\uparrow$ are propagated downwards, using the transposed connection matrix $\WT$ and scaled by the derivative of the neuronal activation function $r'=\varphi'(u)$.
            Here, $\cdot$ denotes the Hadamard product.

            \item[Prospective coding:\quad $\ubrever = u + \taur \dot{u}$]
            Some models use prospectivity to advance neuronal outputs, %
            see \cref{box:preprospective} for more details.
    \end{description}

 \end{multicols}
\end{boxWithCounter}

\begin{figure*}[t]
    \centering
    \includegraphics[width=\textwidth]{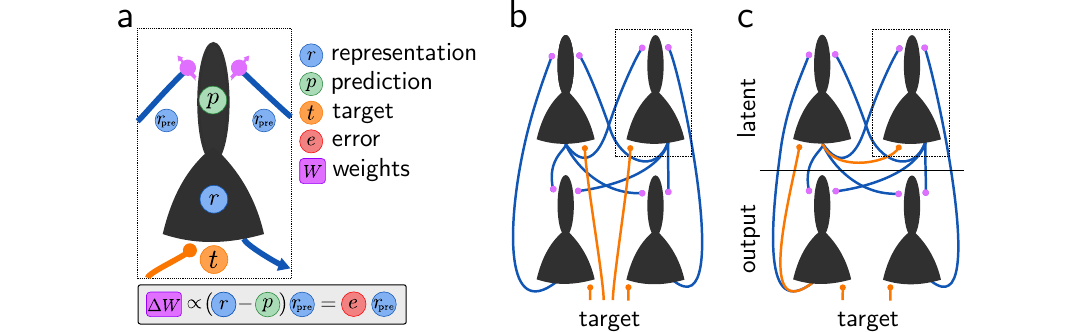}
    \caption[]{\label{fig:model_implicit}
    \textbf{Implicit error representation.}
    \textbf{(a)} \Gls{us} neuron and synapse model \parencite{urbanczik2014learning}.
    Neurons are modeled as having two compartments, with the soma encoding the representation $r$ and the dendrite encoding the prediction $p$ of somatic spiking.
    Soma-targeting teacher synapses nudge the somatic activity towards a target $t$, generating an implicitly represented error signal $e=r-p$.
    Plastic dendrite-targeting synapses are adjusted by a three-factor learning rule to minimize the difference between dendritic prediction and somatic target.
    Here, we also define the color code for our figures, which helps highlight common themes of neuronal coding and synaptic plasticity across all models.
    To this end, the depicted quantities are only symbolic and do not adhere to strict physical units, flexibly representing signals encoded by spikes, rates or membrane potentials.
    \textbf{(b)} Recurrent network with external teachers \parencite{urbanczik2014learning}.
    A target input is provided to all neurons in a recurrently connected population.
    Dendritic synapses adapt to produce predictions matching the targets.
    After learning, the network reproduces the target patterns on its own, in absence of the target input.
    \textbf{(c)} Recurrent network with external and internal teachers \parencite[\glsfmtshort{elise};][]{kriener2024elise}.
    Here, signal transmission delays are actively used for sequence learning.
    Targets are presented only to the output population and propagate from there into the latent population via a sparse, static scaffold that emerges during network development.
    Synapses use the same learning rule as in (b), but here they learn to properly gate delayed activity, allowing the network to learn complex, non-Markovian target sequences.
    }
\end{figure*}

\section{Implicit error representation}
\label{sec:implicit}

In the following, we discuss a set of models based on two-compartment neurons, which provide the prerequisites for error-correcting three-factor plasticity rules.
In all models, the somatic compartment encodes a target signal of some form, and the dendritic compartment forms a prediction of the somatic activity.
This somato-dendritic mismatch acts as an error signal, driving plasticity that minimizes it.

The main differences between the models lie in the nature of the predicted quantity and in how the target signal is generated at the soma.
In the \gls{us} model (\Cref{sec:US}), an external teaching signal drives the soma through static synapses, and the dendrite learns to predict this target activity.
Somato-dendritic consistency learning (\Cref{subsec:somatodendritic_consistency}) instead needs no external teacher: the soma normalizes its own membrane potential, and the mismatch with the unnormalized dendritic prediction drives unsupervised learning of recurring patterns.
\Gls{elise} (\Cref{subsec:elise}) extends the \gls{us} model to \glspl{rnn}, allowing latent neurons to provide teaching signals to each other, in addition to those coming from external targets.

\subsection{Learning by dendritic predictions of somatic spiking}%
\label{sec:US}

As a first example of a learning algorithm where errors are computed locally and appear only implicitly, we consider the \gls{us} model \parencite{urbanczik2014learning}.
Here each neuron is modeled as having one somatic and one dendritic compartment.
These compartments are innervated by soma-targeting and den\-dri\-te-targeting synapses with distinct properties and functional roles.

Soma-targeting synapses are static and carry a teaching signal.
As the somatic compartment is conductively coupled to the dendrite, in the absence of teaching input the somatic potential is fully driven by and will thus quickly approach that of the dendrite.
In the presence of a teaching signal, the voltage of the soma is pulled away from the dendritic potential by the conductance-based teaching synapses.
The learning rule adapts the dendrite-targeting synapses according to $\Delta W \propto \left[\varphi(u) - \varphi(v^*) \right] \rpre$ to minimize the postsynaptic error represented as the difference between the activity of the teacher-nudged soma $\varphi({u})$ and the activity of the dendrite $\varphi(v)$ (with $\rpre$ being the presynaptic firing rate).

In this and other related models, the learning rule needs a small but important correction in the dendritic term to account for the somatic leak.
By setting $v^* = (\gl \El + \gsd v) / (\gl + \gsd)$, the dendritic term recovers the somatic potential in the absence of nudging, thereby allowing the unbiased representation of the teacher-induced error.
Here, $\gl$ and $\gsd$ represent the leak and the somato-dendritic coupling conductances, and $\El$ the leak potential.

This plasticity mechanism can endow recurrently connected networks with capabilities that are widely recognized as fundamental building blocks of cortical computation \parencite{urbanczik2014learning}.
For example, in an auto-associative memory task, different patterns can be imprinted onto a randomly connected recurrent network by having a teacher directly nudge the somata of the relevant neurons (\cref{fig:model_implicit}a).
The \gls{us} rule allows the (recurrent) dendritic weights to adjust until the dendritic prediction matches the somatic potential, explaining away the effect of the teacher.
The stored patterns thus become attractors of the recurrent dynamics and a brief somatic nudge is enough to push the network into the corresponding sustained activity state.

A second example involves the unsupervised learning of a self-organizing topographic map.
Here, the plastic dendritic synapses carry feedforward input from a stimulus layer, while the somatic nudging is supplied by the network itself through balanced short-range excitation and long-range inhibition via interneurons.
Each neuron thus learns to predict neighboring activity while being pushed away from responses of distant neurons, creating a topographic mapping in which correlated stimulus features are mapped to neighboring neurons, reminiscent of cortical feature maps.

\begin{sloppypar}
An extension by \textcite{brea2016prospective} keeps the two-com\-part\-ment \gls{us} neuron model but modifies the temporal properties of its plasticity rule.
The somatic component is now driven by a low-pass-filtered, and hence lagged, presynaptic trace $\rprebar$, resulting in the learning rule $\Delta W \propto \varphi(u)\rprebar - \varphi(v^*)\rpre$.
To reach equilibrium, the dendrite then needs to learn to predict the \emph{expected future} somatic activity.
Importantly, when multiple presynaptic inputs are present, the neuron can learn to predict its own predictions over behavioral time horizons.
Since the somatic potential already contains the dendritic contribution, each newly learned anticipation becomes a target for further anticipation, and the learned activity expands backwards in time.
This bootstrapping effect allows the neuron to learn a temporal prediction horizon that is much longer than the mere plasticity window.
By associating the predicted signal with a reward, neurons can learn to predict a discounted future reward, relating this learning rule to the TD($\lambda$) algorithm from reinforcement learning.
\end{sloppypar}

The ability of a neuron to predict future inputs is referred to as prospective coding.
Here, prospectivity is not an intrinsic property of neuronal dynamics, but arises through a form of synaptic plasticity that is enabled by the neuron's compartmentalization; we will return to a different variant of prospectivity in the context of deep learning in \cref{sec:xle}.

\paragraph{Biology}
The \gls{us} model assumes a two-compartment neuronal structure in which dendritic inputs act as a prediction of somatic firing, while a separate teaching input nudges the somatic activity towards a target.
The discrepancy between dendritic prediction and teacher-nudged somatic representation then drives learning in the dendritic inputs.

The dissociation between somatic and dendritic activity, as required by \gls{us} learning, is well-documented.
\textcite{moore2017dynamics} found that dendritic spikes in freely behaving animals occur several-fold more frequently than somatic spikes, with large sub-threshold dendritic voltage fluctuations.
Further, \textcite{miles1996differences} showed that peri-somatic inhibition can strongly suppress somatic firing while leaving dendritic responses comparatively intact.

A second requirement is that plasticity is jointly gated by the local dendritic potential and by somatic activity. The dependence on local depolarization is well established, with evidence of the sign of plasticity being set by the level of postsynaptic depolarization \parencite{artola1990different, sjoestroem2001rate}.
Regarding the availability of somatic activity, experiments have shown that action potentials propagate back into dendrites where they influence local calcium dynamics and the sign of synaptic plasticity \parencite{sjoestroem2006cooperative, wright2025distinct}.
Further experimental support for error-correcting learning in cortical microcircuits is discussed in \cref{sec:explicit}.
The circumstantial evidence for the \gls{us} model is thus quite diverse, but both the predictive relationship between the two compartments and the specific plasticity mechanism of dendritic synapses have yet to be quantitatively measured.

A strong prediction of the model is that dendritic and somatic activity should co-evolve, simultaneously moving closer to each other and to the target signal -- this is the essence of the somatic nudging principle.
Also, with such convergence a neuron should become increasingly insensitive to removal of the teaching input.
Experimental verification of this long-term trend could thus exploit simultaneous longitudinal recordings in the soma and dendrite \parencite{gillon2024responses}.

\begin{sloppypar}
Considering the \gls{us} model for associative memory learning, there appear to be some correspondences to the architecture supporting such learning in hippocampal \gls{ca3} \parencite{kesner2013process}.
Pyramidal neurons in \gls{ca3} form a recurrently connected network and receive strong mossy fiber input from \gls{dg} \parencite{lisman1999relating}.
Consistent with the functional division between teaching and predictive inputs in \gls{us} learning, mossy fiber input gates plasticity at recurrent \gls{ca3} synapses through local dendritic \gls{nmda} spikes \parencite{brandalise2014mossy, brandalise2016dendritic}.
Plasticity itself depends on the relative timing of the two pathways, yielding \gls{ltp} when \gls{ca3} recurrent input precedes mossy fiber input and \gls{ltd} when the order is reversed.
This is compatible with a modified version of the \gls{us} plasticity rule in which both dendritic and somatic activities leave short memory traces, for example by being low-pass-filtered at the synapse.
As mossy fibers are highly plastic themselves \parencite{nicoll2005synaptic}, this would correspond to teaching inputs in the \gls{us} model that are variable in time.
Importantly however, as predicted by removing the teacher in the \gls{us} model,
inactivation of mossy fiber synapses impairs spatial learning but not subsequent retrieval \parencite{lassalle2000reversible}.
\end{sloppypar}

\subsection{Somato-dendritic consistency learning}
\label{subsec:somatodendritic_consistency}

One core assumption of the original \gls{us} learning rule is that the instantaneous activities produced by the dendrite and the soma are identical up to the difference elicited by a teaching signal available only to the soma.
A more recent model breaks with this assumption, yielding a mechanism for learning to detect patterns in an unsupervised way~\parencite{asabuki2020somatodendritic}.
Here, unlike in other models using \gls{us} learning \parencite{urbanczik2014learning, brea2016prospective, kriener2024elise}, the soma does not receive teaching input.
Instead, the discrepancy between the soma and the dendrite, which is crucial for learning, arises from differences in how the instantaneous firing rates of the soma $\varphi(u)$ and the dendrite  $\varphi(v^*)$ are computed.
As per usual in the dendritic compartment, the instantaneous outputs are computed as $\varphi(v^*)$.
However, the soma applies the (sigmoid) activation to a normalization of its membrane potential.
Specifically, it tracks both the running mean $\mu$ and standard deviation $\sigma$ of its membrane potential and uses these quantities to produce activity according to $\varphi(\hat{u})$ where $\hat{u} = \frac{u - \mu}{\sigma}$.
As this normalization balances $\hat{u}$ around $0$, activity does not saturate and is always in the steepest, most sensitive part of the activation function.

To understand the implications of this adaptation, consider a single neuron whose dendrite receives irregular spiking input through small random weights.
Whenever some of these input spikes coincide, the dendritic potential deflects slightly.
Because the somatic activity is constantly normalized, it is always sensitive even to small deviations, and the soma responds stronger to dendritic input than the dendrites alone would predict.
As in the standard \gls{us} rule, this mismatch triggers plasticity, but since the input spikes are random, none of the dendritic weights is systematically potentiated, and all weights remain small under weight decay.
This changes when noisy inputs are interleaved with irregularly repeated frozen temporal patterns.
Now, the same synapses are repeatedly engaged in the same order, and by adapting the input weights, the dendritic voltage can learn to approximate the somatic one.
As learning progresses, the somatic response and hence its variance grows, which leads to an adaptation of the neuron's activation function and stronger input -- the input from the learned pattern -- is now required to invoke the same output firing frequency.
Thus, the initial noise sensitivity ultimately turns into a high pattern selectivity.

Since the pattern selected by a neuron is arbitrary, diversity of learned patterns is enforced by lateral inhibition, which leads to the formation of clusters of neurons which preferentially respond to diverse structures in an input.
This property can be leveraged to train networks to perform temporal feature segmentation tasks including learning of orientation tunings and detection of temporal communities.

\subsection[\Glsfmtlong{elise}]{\Gls{elise}}%
\label{subsec:elise}
The \gls{elise} model adapts the \gls{us} learning rule for sequence learning in recurrent neural networks~\parencite{kriener2024elise}.
Sequence learning is a difficult task, especially in the case of non-Markovian sequences, where the next state depends not just on the current state, but on some portion of the history of past states (e.g., the sequence \texttt{A B A C}).
Accordingly, for successful learning, networks need to store memories of past states in their dynamics and shape them in a way that is helpful towards producing the desired output sequence.

In \gls{elise}, as in other models, this is achieved via a recurrently connected latent population connected to an output population.
As evidenced by the limited capabilities of reservoir models, only adapting output weights is not efficient for learning complex temporal patterns \parencite{vlachas2020backpropagation}.
To fully harness the \gls{rnn}'s capabilities for learning complex patterns, the \gls{elise} model goes beyond training of only the readout weights, and constructs a learning signal also for the recurrent connections.
Recall that \gls{us} learning requires each synapse to have local access to three quantities: the neuron activity in absence of the teacher (dendritic voltage), the activity in its presence (somatic voltage) and a trace of presynaptic activities.
The challenge in the \gls{rnn} setting is to provide useful targets for latent neurons.
Ideally these should be structured so that temporal information is distributed spatially inside the network, with some neurons containing memories of past states around which the rest of the network can adapt to produce the correct outputs.
To achieve this combination of structure and plasticity, two types of connections are introduced, both of which are associated with delays.
During an initial developmental phase a sparse scaffold of soma-targeting connections (the orange connections in \cref{fig:model_implicit}) is formed.
These connections remain fixed during learning and thus propagate targets from the output into the latent population in a structured way.
Intuitively, the scaffold translates temporal information into spatial information.

During the subsequent learning phase, a dense matrix of dendrite-targeting synapses (green connections in \cref{fig:model_implicit}) adapts so that neurons learn to spontaneously reproduce the target present at the soma.
As in the original \gls{us} model, the errors that drive learning are (implicitly) computed as the difference between target (somatic) and predicted (dendritic) activities.
Once learning has converged, both latent and output neurons are driven onto their targets by dendritic inputs, thus rendering teaching input obsolete.
Accordingly, the network is able to replay the sequence robustly in the absence of teaching input.

The use of dendrites endows \gls{elise} with measurable advantages over other models for biologically plausible recurrent weight learning \parencite{gilra2017predicting, sussillo2009generating, maes2020learning}.
One set of advantages derives from the inherent spatial and temporal locality of learning in each synapse in \gls{elise}.
One way in which related models achieve similar locality is to make error signals computed at the output nearly instantaneously available to synapses in the network \parencite{sussillo2009generating, gilra2017predicting}.

Another approach is to store activities produced by latent neurons in the form of an eligibility trace until errors have propagated back from the output -- see \textcite{murray2019local} and \textcite{bellec2020solution} for rate- and spike-based versions of such eligibility-based approximations to \gls{rtrl}~\parencite{williams1989learning}, as well as \cref{sec:xle} for a more in-depth discussion.
However, both the instantaneous transportation of errors and the storage of complex eligibility traces are difficult to reconcile with biology.
Importantly, both approaches are incompatible with the presence of transmission delays.
In contrast, \gls{elise} shows how delays can help drastically decrease the neuronal real-estate required for learning, allowing networks with as few as 30 latent neurons learn long complex non-Markovian sequences efficiently and robustly.
This compares favorably to the thousands of latent neurons required by models relying on the slowness of neuronal dynamics alone \parencite{sussillo2009generating,gilra2017predicting} or long carefully calibrated transition chains \parencite{maes2020learning} to retain memories of past inputs.

\paragraph{Biology}

As birdsong is a prominent example of pattern learning and replay, \gls{elise} lends itself as a mechanistic model for understanding birdsong learning in zebra finches.
As juveniles, zebra finches learn a single song based on that of their tutor \parencite{bolhuis2006neural, hahnloser2010auditory}.
During learning, activity in pre-motor area \gls{ra} becomes increasingly self-similar \parencite{oelveczky2011changes, okubo2015growth}.
Similarly, in \gls{elise}, learning structures the activity in the latent population until it becomes almost identical across iterative replays.
Furthermore, due to the teaching scaffold, teaching signals in the model are structured from the beginning of learning.
Similarly, it has been shown that songbirds are born with a species-specific template that is refined through listening to the tutor \parencite{konishi1965role, gobes2007birdsong, bolhuis2015birdsong}.
As in \gls{elise}, once learning has converged, the removal of the teacher has no impact on replay.
Likewise, once a bird has crystallized its adult song, the removal of areas encoding the tutor song impairs the recognition of the tutor song but not the song production \parencite{gobes2007birdsong}.

Finally, there seems to be a correspondence in the way \gls{elise} and birdsong learning utilize sources of unstructured random activity to improve learning.
At the onset of learning, the randomly initialized teaching synapses weakly propagate activities.
With learning, connections that contribute to correct replay are singled out and reinforced.
In songbirds, the source of this variability is the \gls{lman}, which projects onto pre-motor area \gls{ra} \parencite{kao2005contributions, aronov2008specialized}.
Where \gls{lman}-driven variation shapes the produced song, subcortical structures are thought to trigger plasticity that subsequently biases \gls{lman} towards the activity patterns that proved useful \parencite{fee2011hypothesis, gadagkar2016dopamine}.
All-in-all, these correspondences between the \gls{elise} model and song learning point to the interplay of structure and randomness as a general requirement for sequence learning systems.
Here, the presence of dendrites plays a crucial
role in separating the structures useful for propagating teaching signals (via the static sparse teacher scaffold) from the initially random somato-dendritic connections that enable flexibility during learning.

\section{Explicit error representation}%
\label{sec:explicit}

In models where errors are represented explicitly, mismatches between expected and observed stimuli are directly accessible as measurable quantities.
In \emph{hierarchical} models of the cortex, these theories are often closely related to deep learning models.
With explicit error representations, network performance can be measured by synapses through the magnitude of local error signals.
This represents an explicit form of credit assignment, as by influencing local errors, synapses can ultimately have an effect on the global network output.

Most of the models in this category are based on a tripartite neuron model:
Inspired by the morphology of cortical pyramidal neurons, the core computational units consist of a prediction dendrite integrating afferent input and therefore encode the currently learned internal representation.
A second compartment locally encodes an error signal and is, depending on the strength of the coupling between the error compartment and the soma, weakly or strongly nudged towards the intended representation.
Loosely speaking, the prediction dendrite encodes what the neuron \emph{thinks} it should do and the soma encodes what the neuron should \emph{actually} do, as dictated by the target output.
This mismatch is picked up and gradually reduced by er\-ror-cor\-rec\-ting learning rules similar to those discussed above.

The differences between the models in this section lie primarily in the type of teaching signal being provided at the error compartment, the nudging strength and the error transport mechanism across hierarchies.
Information representation and learning rules are further shaped by the specific temporal dynamics of each neuron model.

\begin{sloppypar}
Depending on whether models focus on generative (e.g., pattern completion) or discriminative (pattern recognition) properties of cortex, predictions can flow either from primary sensory to higher-order areas or vice versa.
While few models address both pathways simultaneously (however, see~\cite{deperrois2022learning}), it is probably the case that cortex does both.
\end{sloppypar}

Error transport across cortical areas
varies substantially between models:
Models of dendritic error construction (\Cref{subsec:dendritic_error_construction}) do not transmit the error as a separate signal but reconstruct it locally at each area from the mismatch between incoming projections and interneuron predictions, so only the combined signal $r=p+e$ travels between areas.
{\Glspl{enmc}} (\Cref{subsec:error_neuron_MCs}) use a dedicated population of error neurons that propagates errors from area to area alongside a separate representation pathway, extending the compartment-free, two-population circuit of \textcite{whittington2017approximation} that mirrors backpropagation.
\Gls{dfc} broadcasts a centrally computed mismatch signal to all areas at once, approximating a second-order, Gauss-Newton solution to credit assignment.
Bursting models instead multiplex activity and error onto a single output channel, using isolated spikes and bursts to carry each signal separately~\parencite{payeur2021burstdependent,greedy2022singlephase}.
\Gls{le} and its extensions (\glsfmtshort{gle}, \glsfmtshort{vle}) act on the transport mechanism itself, generalizing it to prospective communication, in which neurons signal anticipated future states rather than present ones.
This overcomes the relaxation problem and solves spatio-temporal credit assignment by approximating non-local-in-time future errors, mapping onto either \glspl{dcmc} or \glspl{enmc}.

\subsection{Dendritic error construction through local inhibition}%
\label{subsec:dendritic_error_construction}

Models of dendritic error construction compute the local error directly within a representation neuron's dendritic compartment, rather than receiving it from a dedicated population of error neurons as in \Cref{subsec:error_neuron_MCs}: the same neuron that uses the error for its own plasticity is also the one that constructs it.

\begin{figure*}[t!]
    \centering
    \includegraphics[width=\textwidth]{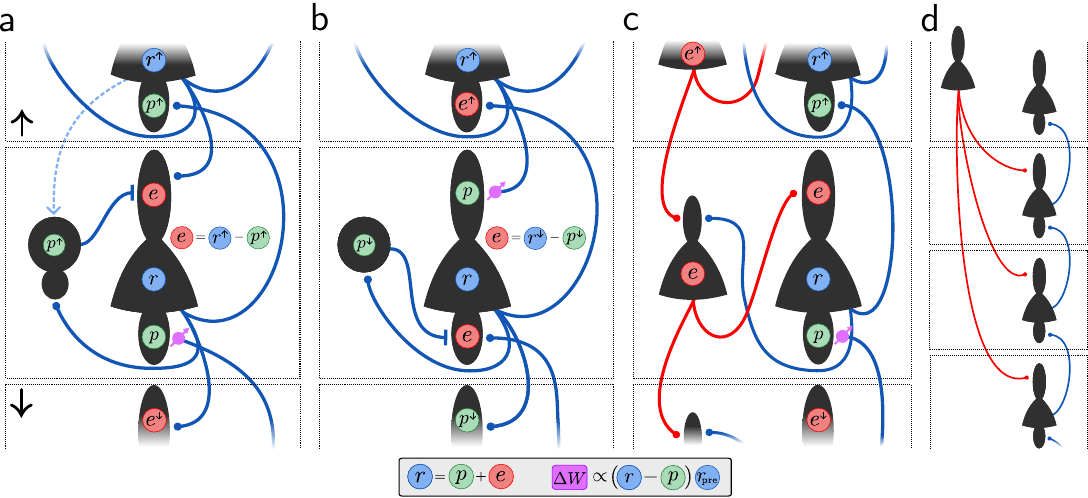}
    \caption[]{\label{fig:model_explicit}
    \textbf{Error representations in cortical microcircuits.}
    In all microcircuit models, pyramidal neurons have a similar structure.
    Dendritic errors $e$ nudge somatic representations $r$ away from dendritic predictions $p$, allowing synapses access to the necessary information in order to instantiate a form of the error-correcting delta rule $\Delta W \propto (r - p)\rpre = e \rpre$.
    \textbf{(a)} In models of dendritic backpropagation~\parencite{sacramento2018dendritic, haider2021latent, max2024learning} and target propagation \parencite{galloni2026cellular},
    inhibitory interneurons learn to copy predictions of the area above $p^{\uparrow}$ through teacher synapses coming from those areas (dashed arrow).
    In the error compartments of pyramidal cells, interneuron activity is subtracted from inter-area projections $r$, forming a local error representation: $e = \alpha (r^{\uparrow} - p^{\uparrow}) = \alpha e^{\uparrow}$.
    The proportionality factor $\alpha = \varphi' \WT$ is determined by the derivative of the top-down neuronal firing rate $\varphi'$ and the top-down synaptic weights $\WT$, yielding an approximation of the \gls{bp} algorithm.
    See \Cref{box:weightTransportProblem} for a discussion of how backward weights can become transposes $\WT$ of the forward weights.
    \textbf{(b)} Dendritic hierarchical predictive coding~\parencite{mikulasch2023where} employs the same error coding and synaptic plasticity mechanism, but applied to an inverted hierarchy of areas.
    The dendritic compartments are effectively swapped with respect to (a), with the basal dendrite now representing the error $e$ and the apical dendrite representing the prediction $p$.
    \textbf{(c)} In dendritic error neuron microcircuits~\parencite{ellenberger2025backpropagation,max2026backpropagation,brandt2026variational}, there are two populations of pyramidal cells: error neurons (left) and representation neurons (right).
    Representation neurons are directly connected, forming a prediction stream across areas (blue, upwards).
    Similarly, error neurons directly project to each other (red, downwards).
    Local connectivity allows error neurons to project their activities onto apical dendritic compartments of representation neurons.
    The factors $\WT$ and $\varphi'$ are separately transported towards error neurons by top-down synapses from other error neurons and horizontal connections from representation neurons, respectively.
    \textbf{(d)} Global error broadcasting models~\parencite{meulemans2021credit, meulemans2022minimizing} do not propagate errors layer-by-layer, but use a global controller in the top layer to adapt local errors in every layer in a way that reduces the top layer error.
    Here, the same error-correcting plasticity rule as in the other models yields an approximation of Gauss-Newton optimization, rather than \gls{bp}.
    }
\end{figure*}

Here, we discuss three rate-based models of error signal construction through local inhibition: \gls{dbp} by \textcite{sacramento2018dendritic},
\gls{dtp} by \textcite{galloni2026cellular}, and \gls{dhpc} by \textcite{mikulasch2023where} (see also \cref{fig:model_explicit}a,b).
\Gls{dbp} and \gls{dtp} operate as discriminative models, whereas \gls{dhpc} is generative, but otherwise they essentially carry out the same error propagation and learning algorithm.
Inverting the area/population hierarchy of \gls{dhpc} yields the architecture of the other two models \parencite[see also][Appendix~C]{max2026backpropagation}.

The idea of introducing separate dendritic compartments to store different quantities (sensory information and higher-order feedback) for deep credit assignment was first proposed in \textcite{guerguiev2017deep}.
Here too, the error was represented only implicitly, in the temporal difference between a neuron's feedforward- and feedback-driven phase.
The models described in this section are modified as to remove the need for such phases and better align with continuously active plasticity.

Motivated by mechanisms of excitatory-inhibitory balance between pyramidal and interneurons in cortex
\parencite{deneve2016efficient}, errors are hypothesized to be computed in dendritic compartments from mismatches between the activities of representation neurons and interneurons.
This makes the error signals locally available to the neuron, without the need for a separate population of error neurons.
The models achieve this by representing pyramidal cells as three-compartment neurons, and using a local population of interneurons.

The working principle is sketched in \Cref{fig:model_explicit}a and b.
When presenting a stimulus in absence of a learning signal, the network settles into a state where both the pyramidal cells and interneurons encode the same predictions.
The horizontal pyramidal-to-interneuron projections are adapted such that the error-representing compartments~$e$ of pyramidal cells have zero activity, i.e., interneuron~$i$ and afferent pyramidal cell activity~$r$ cancel out exactly.
In the models, such matching of activity is referred to as `tight balance'~\parencite{mikulasch2023theory} or the `self-predicting state'~\parencite{sacramento2018dendritic}, and is facilitated by local learning rules.
These adjust lateral, interneuron-targeting connections, based on representations from other (downstream) areas
to match the forward weights.
After establishing such a balanced state, an error signal can be introduced in one area of the network.
Now, the activity of pyramidal neurons in this area encodes a mixture of predictions and error signals, symbolically: $r=p+e$.
This signal is projected to pyramidal cells in other areas (central boxes in~\cref{fig:model_explicit}a,b).
Because interneurons still only encode predictions, the difference between afferent projections from pyramidal cells in other areas and local interneuron activity is precisely the backpropagated error from other areas: $e=r^{\uparrow}-p$ for \gls{dbp} and \gls{dtp}, or $e=r^{\downarrow}-p$ for \gls{dhpc}.
Hence, the compartment voltage~$e$ encodes an explicit error signal.
This mechanism allows error signals to be projected across areas, on top of the prediction signals.

\Gls{dbp} and \gls{dtp} rely on this same mechanism and, since $e=r-p$, arrive at algebraically identical weight updates.
Although named after \gls{tp}, \gls{dtp} reconstructs this top-down error in the same indirect way as \gls{dbp}, from the mismatch between nudged and unnudged, prediction-only activity, rather than via an explicit inverse mapping of targets through the network as in classical \gls{tp}.
Its main differences from \gls{dbp} instead lie in incorporating lateral inhibition, Dale's-law-compliant connectivity and eligibility traces.

The weights connecting representation neurons across areas are adapted to minimize the locally reconstructed error signal, using a delta rule of the form $\Delta W = (r-p) \times  r^\downarrow$ (\gls{dbp}, \gls{dtp}), or  $\Delta W = (r-p) \times  r^\uparrow$ (\gls{dhpc}).
With the appropriate mechanisms described above and under the assumption of small errors, these learning rules have been shown to approximate \gls{bp} \parencite{sacramento2018dendritic}, and have further been related to the error propagation mechanism of Difference Target Propagation~\parencite{lee2015difference,max2024learning}.
One of the underlying assumptions is weight symmetry between forward and backward pathways, also known as the weight transport problem (see \cref{box:weightTransportProblem}).

While simulation results have demonstrated that dendritic error construction can solve non-linear classification problems~\parencite{sacramento2018dendritic,haider2021latent,max2024learning,galloni2026cellular}, these models rely on strong assumptions such as a perfect cancellation of inhibitory and pyramidal activity in absence of external targets (see also the \emph{Biology} section), and extremely slowly varying input (or dedicated relaxation phases otherwise).
This relaxation problem is shared with Equilibrium Propagation~\parencite{scellier2017equilibrium} and requires further mechanisms to plausibly scale to more complex problems \parencite{haider2021latent} (see also \cref{sec:xle}).
It also remains unclear whether dendritic error construction through local inhibition can scale to propagate useful error signals across more than two areas~\parencite{max2025backpropagation}.

\paragraph{Biology}

Such models of dendritic error construction require tight matching of interneuron and pyramidal cell activity.
This fits well with observations of strong correlation between stimulus-evoked excitation and inhibition in cortex~\parencite{wehr2003balanced,okun2008instantaneous}, even at the dendritic level~\parencite{liu2004local}.
The excitatory units are assumed to be pyramidal cells~\parencite[e.g.,~in cortical layers 2/3,][]{mikulasch2023where}, each projecting their predictions to up- and downstream areas (\cref{fig:model_implicit}a,b).
Inhibitory neurons are required to project locally onto the dendrites of pyramidal cells.
As candidates, \gls{sst} interneurons have been discussed in~\textcite{sacramento2018dendritic}, as they preferentially target apical dendrites of pyramidal cells, in line with the model's assumption.
Alternatively, \gls{pv} basket cells have been proposed, due to their ability to quickly inhibit somata and basal dendrites of pyramidal cells with fast spiking~\parencite{mikulasch2023where}.

However, the tight matching condition is rather strict: it requires close correspondence between interneurons and pyramidal cells, both in connectivity and activity.
This does not necessarily need to hold at the level of individual cells and synapses, since populations of interneurons could instead learn to match populations of pyramidal cells, leaving some freedom for individual units, but dendrites still require the average match to hold closely enough for local error signals to be computed.
Given that interneurons non-linearly integrate inputs from different cells~\parencite{cornford2019dendritic}, such a tight matching is still difficult to reconcile with biological evidence.

Models of dendritic error construction have distinct experimental signatures.
Foremost, as discussed above, interneurons should closely match pyramidal cell activity in absence of prediction errors, at least at the population level.
Further, error signals are encoded in either basal~\parencite{mikulasch2021local} or distal apical~\parencite{sacramento2018dendritic,galloni2026cellular} dendrites of pyramidal cells, and somata integrate error signals with predictions from other pyramidal cells.
Detailed experimental validation would thus make it necessary to record simultaneously from dendrites and somata \emph{during learning}, a technique for which only few datasets exist \parencite[e.g.,][]{gillon2024responses}.

\subsection{Error neuron microcircuits}%
\label{subsec:error_neuron_MCs}
\glsreset{enmc}  %

In the models discussed above, pyramidal neurons in the forward pathway are paired with interneurons that can also be considered as being part of the same pathway, and their difference represents the (backpropagated) error; backward propagation thus happens only implicitly, without a dedicated backward pathway.
By contrast, in error neuron microcircuits, errors are represented explicitly and propagated backward by a separate class of pyramidal error neurons that form an explicit backward pathway, an idea first suggested in the two-population predictive coding circuit of \textcite{whittington2017approximation}.
Lateral projections from these error neurons to the dendrites of representation neurons in the forward pathway then allow the same dendritic representation of errors for error-correcting learning as in \Cref{subsec:dendritic_error_construction}, only now inherited directly from an explicit error neuron population rather than computed locally.

More recently, such a symmetric microcircuit was proposed as a substrate for the \gls{gle} framework~\parencite[see \Cref{sec:xle}]{ellenberger2025backpropagation}, and extended with a detailed study in~\parencite{max2025backpropagation}.
This \gls{enmc} model of cortex is illustrated in~\Cref{fig:model_explicit}c.
The \gls{enmc} model improves on the biological plausibility of previous models by considering a more general connectivity than the often-assumed strict hierarchy we have seen until now; it also lifts the strict one-to-one coupling and self-predicting-state matching discussed in \Cref{subsec:dendritic_error_construction}.
It demonstrably approximates \gls{bp} for learning in rate-based networks, and also enables the approximation of \gls{bptt} for temporal credit assignment \parencite{ellenberger2025backpropagation}.

The computational advantage of explicit error neurons is that they allow an even closer correspondence to artificial networks trained with \gls{bp} compared to error calculation in dendrites.
In the \gls{enmc} model, the two opposing pathways of inference vs.~error backpropagation are implemented at the circuit level:
representation neurons (upwards in \cref{fig:model_explicit}c) project directly to representation neurons in other areas; errors are passed (downwards) directly from error neuron to error neuron \textendash{} this differentiates the model from classical~\gls{pc}, where all projections are bidirectional~\parencite{max2025backpropagation}.
As in \gls{pc}, error signals are explicitly coded in the output of error neurons.
These errors are constructed by multiplying afferent errors from higher areas $e^{\uparrow}$, with a gating signal from local prediction units, implementing the \gls{bp} computation $e = r' \cdot \WT e^{\uparrow}$.
Such gating is implemented by modeling error cells with two compartments, multiplying local predictions $r'$ with afferent errors $\WT e^{\uparrow}$.
As in previous models, this error propagation assumes a form of weight transport, for which biologically plausible plasticity mechanisms exist (see \cref{box:weightTransportProblem}).

Similar to the pyramidal cells in dendritic error construction models~\parencite{sacramento2018dendritic,mikulasch2023where}, each representation neuron has three compartments: one compartment accumulating predictions $p$ from representation neurons in other areas; an error-receiving compartment, onto which local error neurons project -- this makes the error $e$ available to the representation neuron; and a somatic compartment encoding the representation $r$, which integrates predictions and errors from its adjacent dendrites.
Learning is facilitated by a variant of the delta rule, in which representation neurons use prediction errors $r-p$ as a learning signal for plasticity $\Delta W \propto (r-p) r^{\downarrow}$.

The \gls{enmc} model addresses several issues of biological plausibility commonly found in theories of learning in cortex.
As opposed to other \gls{pc} models -- \textcite{rao1999predictive,mikulasch2021local,sacramento2018dendritic}, see also \textcite{mikulasch2023where} for a review --
there is no strict one-to-one matching of neurons in different populations; instead, the model can learn with only loose matching between representation and error units, and different sizes of each population.
This flexibility extends to the hierarchy of areas itself, matching the more complex architectures observed in, for instance, the macaque visual cortex~\parencite{markov2014anatomy}.
In \textcite{max2025backpropagation}, the capabilities of \glspl{enmc} have been demonstrated on various learning tasks, which require errors to be propagated across multiple areas.

\paragraph{Biology}
\label{sec:MC Biology}
The \gls{enmc} postulates two populations of pyramidal cells to serve as prediction and error units, located in cortical layers L5 and L2/3 respectively.
Recent physiological measurements suggest that cortical L2/3 pyramidal cells may serve as error coding neurons, with evidence from various cortical areas, such as visual ~\parencite{thomas2024predictions,zmarz2016mismatch,makino2015learning,jordan2020opposing,gillon2024responses}, visuomotor~\parencite{attinger2017visuomotor}, motor~\parencite{ebina2018twophoton}, and auditory cortex~\parencite{obara2023change}.
L5 pyramidal cells acting as prediction units need to integrate afferent across-area predictions with local error signals.
Therefore, a strong correlation between L5 and L2/3 pyramidal cells within the same area should be observed during learning; after learning or during inference, however, L2/3 error units should be silent, and L5 activity fully stimulus-driven.

The \gls{enmc} model has distinct inter-area connectivity:
it posits two separate, opposing streams of predictions and errors, which project separately across areas, but couple to each other within each area (\cref{fig:model_explicit}c; red vs.~blue streams).
Such distinct feed-forward and feedback pathways have been mapped out across cortical areas~\parencite{markov2014anatomy}.
However, projections of explicit error signals across multiple areas have yet to be experimentally confirmed; in fact, an ongoing study by the OpenScope initiative suggests that error signals are only partially projected across higher-order areas~\parencite{westerberg2025sensory}.
This directly challenges non-dendritic hierarchical \gls{pc} models~\parencite{rao1999predictive,whittington2017approximation},
which require errors to be computed and propagated forward at every level of the hierarchy, but is compatible with the \gls{enmc} model, where predictions, not errors, drive feedforward sensory processing.

In the model, neurons have dendritic compartments with dynamics tailored to the computations they need to implement.
Similar to models of dendritic error construction~(\Cref{subsec:dendritic_error_construction}), L5 pyramidal cells represent predictions and errors in separate dendritic compartments, e.g., basal / perisomatic and distal apical dendrites.
This means that error signals are accessible as separate quantities within the neuron, and should be observable through dendrite-specific recordings.
On the other hand, L2/3 pyramidal cells represent error-dependent activity in their somata, with the dendritic compartment acting as a gating / gain-modulating factor.
Potential biophysical implementations of this gating include a non-linear dendrite, gain modulation through the shunting of synaptic conductances, or short-term synaptic plasticity.
Experimentally, the \gls{enmc} model can be distinguished from models of dendritic error construction through such architectural details: instead of tight matching between interneuron and pyramidal cells, there is explicit error coding in L2/3 neurons.

\begin{boxWithCounter}{box:weightTransportProblem}{The weight transport problem in biologically plausible credit assignment}
\label{box:weight transport}
        \begin{multicols}{2}
            The weight transport problem, coined by \textcite{grossberg1987competitive}, identifies a core implausibility of the backpropagation algorithm \parencite{crick1989recent,whittington2019theories,lillicrap2020backpropagation}:
            According to the recursive error backpropagation rule ${e}={r'}\cdot{\WT e}^\uparrow$, feedback pathways need to carry the exact transpose $\WT$ of the forward weights.
            This constraint requires further attention as no mechanism in biology is known that can directly copy weight information between synapses.
            It affects all algorithms for biologically plausible \gls{bp} presented here.

            Algorithms from the \gls{fa} family \parencite{lillicrap2016random, nokland2016direct} circumvent this problem by replacing $\WT$ by random fixed feedback matrix $B$.
            Despite the observation that forward weights tend to align with the random feedback weights, several studies have demonstrated that \gls{fa} alone performs poorly in deeper networks and when tackling more challenging tasks~\parencite{bartunov2018assessing,moskovitz2018feedback}.

            To address this issue, several weight symmetrization algorithms have been proposed.
            \textcite{liao2016important} have shown that it is  the sign of feedback weights rather than their magnitude that needs to be aligned for meaningful error transport.
            The approach proposed by~\textcite{kolen1994backpropagation} and mapped to a microcircuit structure in \textcite{akrout2019deep} employs weight decay on forward and backward synapses to forget their initial state while updating both pathways with symmetric weight updates.
            However, this relegates the weight transport problem to a weight update transport problem, and it remains unclear how these symmetric updates can be achieved given the omnipresent temporal noise and morphological variability in neuronal substrates.

            More recent algorithms try to infer information about a synapse's reciprocal counterpart from neuronal activity.
            Both \glsfmtlong{pal} \parencite[\glsfmtshort{pal};][]{max2024learning} and its spiking counterpart \glsfmtlong{sal} \parencite[\glsfmtshort{sal};][]{gierlich2026spikebased} exploit noise to estimate the effect of distant weights.
            Other methods such as regression discontinuity design~\parencite{guerguiev2019spikebased} and spike-timing-dependent weight inference~\parencite{ahmad2020overcoming} propose to infer the strength of forward weights through a specific stimulation protocol for the corresponding pre- and postsynaptic neurons.
        \end{multicols}
\end{boxWithCounter}

\subsection{Global error broadcasting}
\label{DFC}

Global error broadcasting was first explored in the \gls{ml} context through direct feedback alignment~\parencite{nokland2016direct}, an extension of feedback alignment~\parencite{lillicrap2016random} that projects the network's output error directly to every layer via a fixed random matrix, instead going sequentially through every layer.
Taking inspiration from control theory, \textcite{meulemans2021credit} cast this broadcasting principle as a genuine dynamical system, the \gls{dfc} framework, which approximates a second-order solution to the credit assignment problem and, as with the models already discussed, maps naturally onto a multi-compartment dendritic neuron model, as detailed below.
They suggest
a \gls{pid} feedback controller that computes a single mismatch signal (equivalent to the output error in \gls{bp}), which is then directly projected to all areas in the network at once.

On the neuronal level, \gls{dfc} is largely identical to the pyramidal neuron model of~\parencite{sacramento2018dendritic,haider2021latent,mikulasch2023where,max2025backpropagation}:
each neuron has three compartments: a feedforward (prediction), feedback (control signal) and somatic compartment (see \cref{fig:model_explicit}c).
The soma additively integrates predictions~$p$ and control signals~$e$ into the neuron's activity~$r$.
Hence, errors are explicitly represented, both in form of the feedback controller, and in the feedback compartments of each neuron.
To update feedforward weights, the same error-correcting learning rule as in \textcite{sacramento2018dendritic} is used, such that $\Delta W \propto (r - p) r^{\downarrow}$.
This approximates Gauss-Newton optimization for a flexible range of feedback weights.
Separately, \gls{dfc} proposes an additional noise-based learning rule to adapt these feedback weights during training.

\Gls{dfc} offers two further advantages over previous models, in terms of biological plausibility.
Usual bio-plausible approaches of implementing \gls{bp} or \gls{pc} carry over the notion that errors need to be propagated through all areas~\parencite[see \cref{fig:model_explicit}]{rao1999predictive,sacramento2018dendritic,mikulasch2023where}, which, in their original formulation, require symmetric connectivity, both in terms of size and distance of projections.
This can, however, be relaxed, as shown in \textcite{max2026backpropagation}, especially by combining such models with a learned weight-alignment mechanism~\parencite[][, see also \cref{box:weightTransportProblem}]{max2024learning,gierlich2025weight}.
Such connectivity, whether symmetric or learned, nonetheless contradicts what is observed in visual cortex, where feedback projections do not mirror feedforward ones~\parencite{markov2014anatomy}.
\Gls{dfc} instead requires no such connectivity by construction, allowing for direct feedback connections to all lower areas and thus providing a better match for the broadly connected feedback patterns observed in cortex.
Combined with a microcircuit motif incorporating local inhibitory interneurons, \gls{dfc} no longer requires a dedicated error compartment~\parencite{rossbroich2023disinhibitory}.

Further, \textcite{aceituno2023learning}~have demonstrated that \gls{dfc} is also applicable to spiking models.
However, the authors criticize the biological plausibility of the underlying \gls{us} learning rule.
They argue that it is unclear how a neuron may compute the difference of activity between its compartments, and transfer the result to the correct synapse.
Instead of using differences between activities in neuronal compartments, \citeauthor{aceituno2023learning}~propose to  use the \emph{temporal} change in neuronal activity as a learning signal.
However, this comes at the cost of slow learning, and it remains to be demonstrated whether their temporal Hebbian learning rule can scale to train large spiking networks.

A follow-up study extends \gls{dfc} to \emph{strong} feedback control, in which the control signal has a strong influence on the activity of representation neurons: $r= p + e$, with $e \gg p$~\parencite{meulemans2022minimizing}.
This is opposed to usual derivations that relate plasticity rules to gradient descent -- effectively all other models in this section --
which only approximate the loss gradient as the feedback vanishes, i.e., for $e \ll p$, also called the weak nudging limit~\parencite{xie2003equivalence,scellier2017equilibrium,haider2021latent,senn2024neuronal,sacramento2018dendritic,max2025backpropagation}.
In the strong nudging case, learning is instead reframed as minimizing the amount of feedback control needed to reach the target, which allows the simultaneous learning of forward and feedback weights without relying on the weak-nudging assumption.

\paragraph{Biology}
As noted above, \gls{dfc}'s three-compartment architecture and learning rule already coincide with those described in \Cref{subsec:dendritic_error_construction}.
The extension of \gls{dfc} to strong feedback control, as described above, also enhances its biological plausibility:
an analysis~\parencite{aceituno2024challenging} of in vitro and in vivo recordings suggests that cortical neurons are strongly influenced by targets during learning, aligning with the strong control hypothesis.
A stronger control signal is also more robust to temporal noise, an arguably intrinsic property of biological substrates.

\subsection[Prospective coding for backpropagation\texorpdfstring{\\}{ }through space and time]{Prospective coding for backpropagation through space and time}\label{sec:xle}

If learning requires settling to an equilibrium, models face a fundamental obstacle towards biological plausibility -- the so-called relaxation problem.
When representations and errors are misaligned in time, error-correcting plasticity can guide synapses in the wrong direction and learning can be substantially impaired.
If neurons are only capable of laggy signal transmission due to the low-pass-filtering effect of membrane integration, such misalignment is inevitable, especially when both predictions and errors need to be transmitted across multiple stages of the cortical hierarchy.
Models such as~\parencite{rao1999predictive,sacramento2018dendritic,mikulasch2023where,scellier2017equilibrium} solve the problem by introducing explicit relaxation times after each sensory input, during which the input needs to remain unchanged.
These can quickly become biologically implausible, considering the speed with which sensory information can change.

\Glsfmtlong{le} \parencite[\glsfmtshort{le};][]{haider2021latent} proposes a solution to the relaxation problem by harnessing prospective coding at the level of individual neurons.
Prospective neurons fire in anticipation of their future input (on a time scale of $\taur$) by modulating their output not only as a function of their membrane potential, but also of its temporal derivative: $\varphi(\breve u) = \varphi(u + \taur \dot u)$.

\begin{sloppypar}
\gls{le} network dynamics are derived from first principles.
Neuron dynamics follow from conservation of a global energy function that accumulates neuron-local energies
across the entire network.
Each of these relate to neuron-local errors $e = \breve u - W \rpre$ that effectively represent a difference between what a neuron does and what its presynaptic partners think it should do.
With this, the two main differences from classical \gls{pc} become apparent: there, neuron dynamics follow gradient descent on an energy (rather than conservation) and errors are defined on instantaneous (as opposed to prospective) membrane potentials.
\end{sloppypar}

Since the model does both predictive coding (acquired through synaptic plasticity across the modeled hierarchy) and prospective coding (as an intrinsic property of neuronal dynamics), the two effects can easily be confused.
For clarity, the difference is summarized in \cref{box:preprospective}.

\begin{boxWithCounter}{box:preprospective}{Predictive vs. prospective coding}
        \begin{multicols}{2}
            \begin{center}
                \bf{Predictive coding}
            \end{center}
            Predictive coding encompasses a wide range of models for cortical hierarchies in which higher-level areas try to predict the states of lower-level areas.
            Unpredictable or novel input leads to neuronal activity that is propagated upwards as an error.
            Error neurons encode the mismatch between true and predicted activities and gate the information flow throughout the network.
	        During learning, connections between neurons are updated such that this mismatch is minimized.
            Thus, after learning has converged, error neuron activity is greatly diminished.
            Such models are often used for explaining the representation of static visual stimuli in cortex.
            \begin{center}
                \bf{Prospective coding}
            \end{center}
             Prospective coding refers to a neuron's ability to anticipate its own near-future state.
             This can either be achieved through synaptic plasticity or as an intrinsic property of neuronal dynamics.
             In the latter case, this anticipation is due to various ionic currents that couple negatively into the membrane equation.
             This can be simplified by considering an activation that depends on both the membrane potential and its derivative: $r = \varphi(\breve u) = \varphi (u + \tau \dot u)$.
             The prospective time constant $\tau$ controls the time scale of the anticipation.
             Different combinations of membrane and prospective time constants allow neurons to attend to different time frames in both their past and their future.
        \end{multicols}
\end{boxWithCounter}

The network architecture that implements the sought neuronal dynamics was originally mapped to the \gls{dcmc} model~(\cref{subsec:dendritic_error_construction}), with apical dendrites constructing errors from differences in pyramidal and interneuron activities.
It can, however, also be interpreted in terms of the \gls{enmc} model (\Cref{subsec:error_neuron_MCs}), with errors computed by distinct population of error neurons.
As a consequence of prospectivity, the transmitted errors and predictions are virtually instantaneous, and thus impervious to quickly changing sensory input.
\gls{le} can thus achieve the same performance as \gls{bp} in \glspl{ann}, but with fully local, biologically plausible plasticity and without the need for relaxation phases.

The specific choice of the prospective time scale constitutes the main limitation of \gls{le}: because the prospective output exactly cancels the neuron's membrane integration, no information about the past is retained, precluding any task that requires memory.
Many sensory processing tasks, however, require memory well beyond a single neuron's membrane time constant (a few tens of milliseconds) -- for instance, on the order of hundreds of milliseconds for the fast processing of movement or sound.
To extend \gls{le} towards temporal processing, \glsfmtlong{gle} \parencite[\glsfmtshort{gle};][]{ellenberger2025backpropagation} decouples the time constants of the prospective output and the membrane filtering, thus enabling it to attend to different time horizons lying either in its past (retrospective coding, $\taur < \taum$) or in its future (prospective coding, $\taur > \taum$).

Memory alone, however, does not solve temporal credit assignment, which requires computing gradients through time, much as \gls{bp} does across space.
\gls{bptt} generalizes \gls{bp} to temporal tasks by unrolling the recurrent computation over time into an equivalent feedforward network with shared weights, on which \gls{bp} is then applied directly.
In practice, this requires storing and replaying each neuron's past activations in reverse.
This anticausal mechanism is obviously nonbiological, and hence much more difficult to reconcile with brain dynamics~\parencite{lillicrap2019backpropagation}.
The equivalent algorithm for physical, time-continuous systems is the \gls{am} from constrained optimization \parencite{kelley1960gradient,todorov2006optimal,chachuat2007nonlinear}.
Just like \gls{bptt}, the \gls{am} requires either perfectly predicting future states, or running time backwards at certain intervals.

The causal equivalent to \gls{bptt} is \gls{rtrl} \parencite{williams1989learning}.
This allows credit assignment to be performed online, but only at the cost of a substantial memory overhead, and by sacrificing spatial locality.
Many approximations of \gls{rtrl} have thus been proposed, such as \gls{rflo} \parencite{murray2019local}, e-prop \parencite{bellec2020solution}, and \gls{ostl} \parencite{bohnstingl2023online}, which radically truncate the gradients to restore locality.
\gls{gle} takes a fundamentally different approach, directly approximating future errors using the prospective coding mechanism discussed above.

\begin{figure*}[t]
    \centering
    \includegraphics[width=\textwidth]{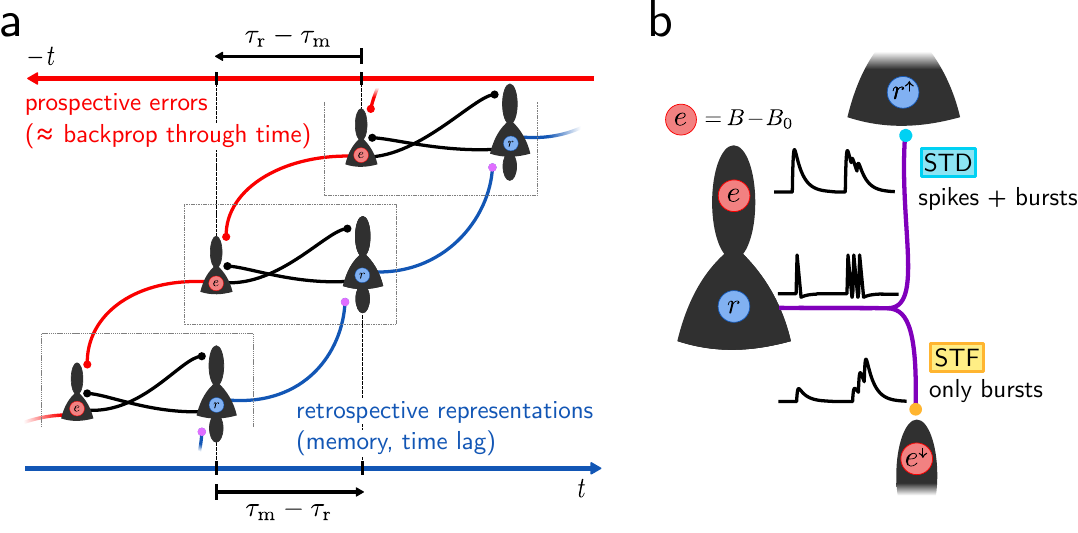}
    \caption[]{\label{fig:burst_prospectivity}
    \textbf{Beyond instantaneous rates: prospective coding and spike/burst multiplexing.}
    \textbf{(a)} Prospective error propagation in \gls{gle}. Representation neurons transmit retrospective, lagged information forward through the network (blue), with each layer inducing a lag on the order of $\taum - \taur$.
    Conversely, error neurons propagate prospective, temporally advanced errors backward (red), effectively looking ahead by $\taur - \taum$.
    Together, these opposite temporal shifts realign errors with the pre-synaptic rates that caused them, approximating \gls{bptt} without explicitly reversing time.
    \textbf{(b)} Multiplexing of spikes and bursts in \gls{burstprop}. Somatic compartments tracking representations emit two types of events: single spikes and bursts.
    The probability that an emitted spike turns into a burst is proportional to the apical activity reflecting an error.
    Forward-propagated activities are filtered post-synaptically via \gls{std} for events, leaving little differentiation between bursts and spikes.
    Backpropagated activities are filtered via \gls{stf}, leaving only bursts to drive apical activity.
    To obtain the error that drives learning, the difference between the bursts ($B$) and a baseline ($B_0$) is computed.
    This scheme captures the core common ideas underlying the different \gls{burstprop} algorithms.
    }
\end{figure*}

In \gls{gle}, the stationarity condition on the energy couples representation and error neurons into dynamical mirrors of each other: by swapping retrospective and prospective time constants, error neurons
invert the temporal lag of their associated representation neurons,
generating errors that are temporally aligned with representations (see \cref{fig:burst_prospectivity}a).
The resulting microcircuit represents the original instantiation of the \gls{enmc} model (\Cref{subsec:error_neuron_MCs}).

The learning rule in \gls{gle} combines this temporally processed error signal with the presynaptic rate, yielding the same delta rule as in \Cref{subsec:dendritic_error_construction}: $\Delta W \propto e \times r^\downarrow$.
For this, representations and (prospective) errors are stored in distinct dendritic compartments, similarly to previously discussed models, and computed in microcircuits as depicted in \Cref{fig:model_explicit}c.

Overall, \gls{gle} approximates the \gls{am}, making it a biologically plausible implementation of \gls{bptt}.
Instead of the causality-violating discounted future error required by the \gls{am},
\gls{gle} error neurons approximate this quantity through prospective coding, which generates the correct temporal shifts for all frequency components of the error signal, at the cost of distorting their amplitude (which is the inevitable price of estimating, rather than knowing, the future).
In practice, these approximate errors work well even in deeper architectures, and \gls{gle} achieves competitive performance on temporal tasks with fully online, local learning.

However, this amplitude offset can also be addressed through appropriate learning of the feedback weights.
By modulating their weight with the average value of the error amplitude distortion, feedback synapses can correct for this effect and in turn improve the learning of the forward synapses.
Such a feedback learning mechanism can be derived within the \gls{vle} framework~\parencite{brandt2026variational}.
Similarly to its predecessors, \gls{vle} derives neuro-syn\-ap\-tic dynamics from an energy function that subsumes all network errors, but instead of assuming energy conservation, it follows a variational approach that minimizes the network's integrated energy over time.
This approach ultimately allows the identification of learning rules for the backward weights $B$ that compensate for mismatches between \gls{gle} and the \gls{am}, aligning the prospective errors $\breve e$ to their future discounted values $\tilde e$: $\Delta B \propto (We - B\breve{\hat e}) \breve{\hat e}$, where $\hat \cdot$ denotes the inverse of the future discount -- a causal operation that can be easily computed.
As in previous models, this assumes a form of weight transport, for which biologically plausible plasticity mechanisms exist (see \cref{box:weightTransportProblem}).
The ability of backward synapses to individually adapt to the prospective properties of their presynaptic error neurons allows \gls{vle} to improve both the stability and the convergence speed of temporal learning in \gls{enmc} networks.

\paragraph{Biology}
The microcircuits used in the latent equilibrium models follow the architectures from \cref{subsec:dendritic_error_construction} and \cref{subsec:error_neuron_MCs} and use the same neuron morphologies.
Here we focus on the ability of neurons to react prospectively to their inputs.
Despite being rarely used in computational models, such prospectivity has been widely observed in biology, from milliseconds \parencite{kondgen2008dynamical} to much longer time scales \parencite{pozzorini2013temporal,lundstrom2008fractional}.
This phenomenon appears naturally as a consequence of negative feedback to the membrane, for example through spike-frequency adaptation or as a component of the Hodgkin-Huxley spike generation mechanism itself \parencite{fuhrmann2002spike, brandt2024prospective}.
Similarly, prospectivity on the scale of the membrane time constant can be observed in populations of noisy spiking neurons without any additional mechanism other than the spike reset itself \parencite{plesser2000escape}.

\subsection{Bursting models}

Standard \gls{ann} architectures consider point neurons capable of transmitting both representations and errors, as needed by the \gls{bp} training algorithm.
In contrast, when modeling cortical neurons -- even compartmentalized ones able to store both errors and representations -- it is often assumed that they can only transmit one type of signal via the somatic output.
This constraint has created the necessity for the microcircuits described thus far (see section \Cref{subsec:error_neuron_MCs}). 
However, considering a more nuanced view of somatic activity can lift this constraint. 
By using bursts and spikes as separate aspects of the neural code, neurons can multiplex their outgoing information, allowing for simultaneous transmission of errors and representations.

The \gls{burstprop} algorithm~\parencite{payeur2021burstdependent,stuck2025burstdependent} and its extension towards \ifglsused{bccn}{\gls{bccn}~\parencite{greedy2022singlephase}}{\glsfmtlong{bccn}~\parencite[\glsfmtshort{bccn},][]{greedy2022singlephase}} utilize bursting mechanisms to enable parallel and independent communication of errors that then drive plasticity.
The model considers  hierarchically organized and compartmentalized neurons which can produce two types of \enquote{spiking events}, single (sodium) spikes and bursts.
While the somatic activity (representation) controls the rate of events the apical activity (errors) controls the probability that an emitted spike turns into a burst. 
Intuitively, as these quantities are mixed in the axonal channel, the post-synaptic neurons need to filter out the relevant signal.
Whereas higher-level somas need to receive all of these representation-encoding events -- i.e., both spikes and bursts -- without differentiating between them, lower level apical compartments want to filter out the errors represented by the bursts.
As shown by \textcite{naud2018sparse}, such a separation of information can be achieved through short-term synaptic plasticity (\cref{fig:burst_prospectivity}b).
Soma-targeting synapses with \gls{std} suppress the effect on quickly repeated spikes (bursts), causing bursts and single spikes to have similar effects on the postsynaptic neuron.
In contrast, apical-targeting synapses with \gls{stf} have the opposite effect: individual spikes have virtually no effect, but repeated spikes trigger a postsynaptic response, such that effectively only bursts are transmitted.
Learning relies on dendritic information segregation similar to the \gls{mc} models described above: weight changes are proportional to a pre-synaptic eligibility trace modulated by a postsynaptic error; the latter is encoded in apical compartments and transmitted through bursts.

These general mechanisms for simultaneous transmission of representations and errors have been cast into a range of computationally performant bursting models.
In the original implementation~\parencite{payeur2021burstdependent}, the probability of a burst is computed as a sigmoid function of the apical activity such that in the absence of top-down input, i.e., with vanishing potential in the apical dendrite,
the neuron produces bursts with some baseline probability $P\_B$.
Learning occurs when the neuron produces an event in the presence of apical activity, resembling a burst-dependent delta rule reminiscent of previous models: $\Delta W \propto (B - P\_B S) E$, where $B$ represents the number of bursts in this context.
Here, the error is represented by the number of bursts $\hat B$ that exceeds the baseline $P\_B S$ present in the neuronal output $S$, while an eligibility trace $E$ forms a running average over the presynaptic input.
If the neuron produces a burst, incoming weights are strengthened in proportion to presynaptic activity, while isolated spikes cause synaptic depression.

Note that even in the absence of an active teacher, so when all apical dendrites should encode zero error, their baseline burst rates are nonzero.
As upstream apical compartments receive bursts from downstream neurons, they move away from their baseline burst rates, causing the representation of spurious errors.
This makes learning phases necessary to determine the baseline burst ratios for each apical dendrite in order to cancel out these spurious errors.

\begin{sloppypar}
To address this issue, the signed version~\parencite{stuck2025burstdependent} introduces two types of bursts.
Long (positive) bursts are produced for an excited apical dendrite $v\geq 0$ and short (negative) bursts are produced if it's inhibited $v<0$ (here, the baseline apical potential is assumed to be at 0).
Rather than having to compare to an average bursting ratio, synapses now receive signed errors, allowing them to directly implement error correction by comparing the rates of positive $B^+$ and negative $B^-$ bursts: $\Delta W \propto (B^+ - B^-) E$.
After successfully learning a task or in absence of a teacher, all apical potentials are zero (no errors) so no bursts are produced, and plasticity does not occur.
This increases the stability of the model substantially, allowing it to learn more complex classification tasks.
\end{sloppypar}

In a further extension of burstprop, \textcite{greedy2022singlephase} take inspiration from~\textcite{sacramento2018dendritic} by modeling another top-down connection:
akin to forward connections, these synapses also implement \gls{std}, thus reacting identically to bursts and individual spikes.
Their weights are set (or learned) to be proportional to the baseline burst probability $P\_B$, such that, on average, they transmit the baseline burst rate $B = P\_B S$.
The difference between the two top-down signals, i.e., between the actual and the baseline burst rates, thus represents an unbiased version of the error.
This difference is computed in the apical dendrite of neurons in lower areas, similarly to how apical errors are computed in \citeauthor{sacramento2018dendritic} from the difference between backward-propagated signals from pyramidal cells and their mirror interneurons.
This makes a baseline correction no longer necessary, allowing continuous learning without phases.

The above studies use different levels of abstraction in their simulations.
While the full spiking model is simulated as described above, \textcite{payeur2021burstdependent,greedy2022singlephase} also use population-averaged rate approximations that also replace the \gls{std}/\gls{stf} gating mechanism by direct transmission of the required signals.
This turns all activities into continuous signals and removes various temporal distortions, which greatly stabilizes learning.
Therefore, while the spiking models were limited to relatively small tasks such as logical XOR~\parencite{payeur2021burstdependent} or low-dimensional regression~\parencite{greedy2022singlephase}, the rate-based models allowed scaling to much more difficult image classification tasks.

\paragraph{Biology}
The neuron model in \gls{burstprop}, like the representation neurons in the models discussed above (see \Cref{sec:MC Biology}), can be mapped onto cortical Layer 5 pyramidal neurons.
These are well documented to intrinsically produce bursting activity, with coincident apical and somatic activation transforming otherwise single spikes into a burst of activity~\parencite{larkum1999new,milojkovic2004burst,shai2015physiology}.
In-vivo results suggest that a multiplexing of spikes and burst is used to convey different information over the same channels~\parencite{naud2024fast}.
In particular, the bursting fraction mimicked detection errors, indicating its role as encoding an error. 
Further, it was shown that dendrite-driven bursting determines synaptic plasticity at the respective dendrites~\parencite{kampa2006requirement, froemke2006contribution}.
On the other hand, as acknowledged by \textcite{stuck2025burstdependent}, the signed bursts are not well covered by experimental evidence, but may be relevant for neuromorphic implementations.

On the postsynaptic side, which is responsible for filtering the relevant quantities, it has been shown that
\gls{std} and \gls{stf} can act as a de-multiplexer of bursts and regular spikes, allowing to distinguish the bottom-up and top-down communication~\parencite{naud2018sparse}.
Additionally, there is recent experimental evidence for connections between suitable neurons with these properties~\parencite{lee2013disinhibitory,kinnischtzke2013motor,petrof2015properties,zolnik2020layer,naskar2021celltypespecific,martinetti2021shortterm}.
As in other hierarchical models, \gls{burstprop} requires symmetry between forward- and backward-projecting synapses -- see \cref{box:weightTransportProblem} for a brief discussion of bio-plausible weight transport.

\section{Uncertainty representation}\label{sec:uncertainty}

The environments to which organisms have to adapt are inherently highly variable.
Taking into account and internally modeling this variability is a necessary component of adaptability.
The brain must hence take uncertainty into account when updating its internal models of the world.
This implies that it is not sufficient for cortical microcircuits to correct for prediction errors and account for unexpected sensorial experiences, but these errors must be modulated in some way by the uncertainty of the prediction.

This adds a toll to the amount of information that neurons need to process, and such modulation would not be realizable in a biologically plausible way without the aid of dendritic compartments to collect and help to locally process all these different pieces of information (predictions, errors, and uncertainties) during learning.

\begin{figure}[t]
    \centering
    \includegraphics[width=\columnwidth]{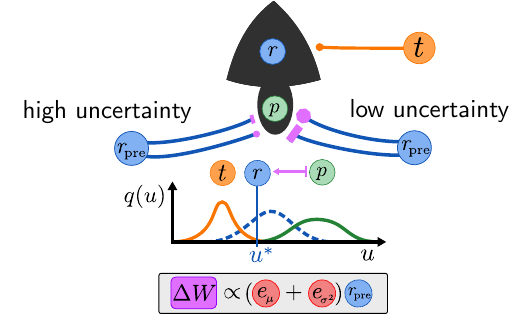}
    \caption[]{\label{fig:bayesian_dendrites}
    \textbf{Dendritic predictive plasticity for reliability-sensitive target learning.} 
    Input from a teacher shifts the somatic distribution away from the one predicted by the dendrite. 
    Learning adjusts dendrite-targeting synapses based on errors $e_\mu$ and $e_{\sigma^2}$ computed from the predicted distribution and samples $u^*$ drawn from the teacher-nudged somatic distribution.
    Synapses carrying low-uncertainty inputs are strengthened over ones carrying more uncertain information.
    The plot (bottom) indicates an example of the probability $q(u)$ of observing a given membrane potential under the teacher (orange), teacher-nudged (blue dashed) and predicted (green) distributions.
    }
\end{figure}

\subsection{Dendritic predictive plasticity performs error correction and reliability matching}

A recent Bayesian view on the dynamics of con\-duc\-tance-based neurons and synapses suggests that they are naturally equipped to perform optimal uncertainty-weight\-ed information integration ~\parencite{jordan2024conductancebased}.
This framework can be seen as an extension of \gls{us} learning whereby the dendrite tries to match a full distribution represented in the soma rather than only an instantaneous value of the somatic potential.
Synapses thus not only learn to reproduce a somatic target activity as in the models presented above, but they also adjust synaptic weights to achieve some target variance in the somatic potential (\cref{fig:bayesian_dendrites}).

Afferent inputs from different sources and associated with different amounts of uncertainty are coupled conductively to the dendrite.
This conductive coupling naturally performs a weighted averaging of inputs, reminiscent of Bayesian opinion pooling. 
In this view, the dendritic dynamics can be interpreted as predicting a posterior distribution over the somatic potential given pre-synaptic inputs.

In the absence of a teaching signal, the somatic membrane potential $u$ will sample from this distribution. 
Given teaching input, this (implicit) distribution will be nudged towards the distribution of the teacher.
Accordingly, these nudged-activity samples $u^*$ can be interpreted as somatic targets and can be used to compute the implicit error signal both for the mean ($e_\mu$) and the variance ($e_{\sigma^2}$) of the predicted distribution.
Dendrite-targeting weights are adjusted to decrease this implicit error, thereby moving the predicted distribution towards the target distribution. 
Furthermore, as afferents with high uncertainty contribute less to producing a correct prediction, they are weakened, thus allowing other projections with higher reliability to gain more influence.

Formally, learning in this model corresponds to reducing the Kullback-Leibler divergence between a target distribution and the currently predicted somatic distribution.
More specifically, plasticity implements gradient ascent on the log posterior somatic probability of samples $u^*$ from the target probability distribution $q(u^*)$: $\Delta W \propto \frac{\partial}{\partial W} \log q(u^*) \propto (e_\mu + e_{\sigma^2}) \rpre$, which gradually reduces the error in both sample mean $\mu$ and variance $\sigma^2$.
Note that the model described in \textcite{jordan2024conductancebased} considers the case where the soma is clamped to the teaching input, thus yielding a direct comparison between a sample from the target and the predicted distribution.
A relaxation of this assumption, whereby the prediction is compared to a sample from the target-nudged distribution (in keeping with learning in the \gls{us} model), seems feasible, while also increasing biological plausibility.

\paragraph{Biology}
\begin{sloppypar}
The model makes the prediction that, after learning, the soma will respond to reliable information with a decrease in membrane potential variability.
One circumstantial piece of evidence is the well-documented wide\-spread quenching of the trial-to-trial variability to a repeatedly presented stimulus \parencite{churchland2010stimulus}.
This reduction in variability can be interpreted as sensory input constraining the set of possible neuronal states, with a more consistent neuronal response reflecting greater certainty about the sensory input.
More recently it was found that stimulus onset in the within-trial setting was associated with a decrease in membrane potential variability proportional to the perceptual reliability in cortical neurons \parencite{huenerbein2025increased}.
Here neuronal spiking responses were recorded in posterior parietal cortex in mice performing an audio-visual change detection task.
The inferred membrane potential variability was found to be lower in trials containing more perceptually reliable multi-modal cues, thereby lending the model direct experimental support.
\end{sloppypar}

\begin{figure*}[t]
    \centering
    \includegraphics[width=\textwidth]{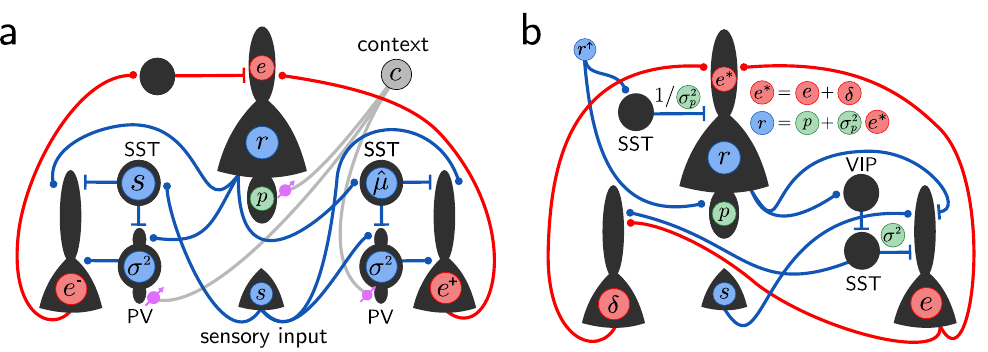}
    \caption[]{\label{fig:uncertainty}
    \textbf{Uncertainty weighting in multi-compartment neurons.}
    \textbf{(a)}~Negative (left) and positive (right) prediction error circuit consisting of three cell types: layer 2/3 pyramidal cells, \gls{sst} interneurons and \gls{pv} interneurons.
    \gls{sst} cells represent the prediction of the stimulus mean in the positive circuit and the stimulus itself in the negative circuit, whereas \gls{pv} cells learn to represent its variance from context.
    Error neurons first form error estimates $(s-\hat\mu)^\pm$ in their dendrites, which are then modulated by divisive inhibition from \gls{pv} cells to form the \glspl{upe} $e^\pm = (s - \hat\mu)^\pm / \sigma^2$.
    These modulated errors then innervate the apical dendrite of representation neurons and drive learning at its basal synapses.
    \textbf{(b)}~A circuit for dynamic uncertainty estimation.
    Following the same principle as in (a) (but with a different assignment of interneuron identities) error neuron circuits produce \glspl{upe} $e$.
    In parallel, second-order errors $\delta=\frac1{\sigma^2}-e^2$ are computed by dedicated second-order error neurons.
    Both $e$ and $\delta$ are integrated into the apical tree of representation neurons.
    Finally, the higher-level (prior) variance $\sigma_p$ multiplicatively controls the gain of apical-to-soma integration.
    }
\end{figure*}

\subsection{Uncertainty-modulated Prediction Errors}\label{sec:upe}

\begin{sloppypar}
While \textcite{jordan2024conductancebased} describe re\-li\-a\-bi\-li\-ty-weight\-ed integration of multiple dendritic input pathways within individual neurons, \textcite{wilmes2025uncertaintymodulated} addresses a complementary problem: how can uncertainty associated with a prediction modulate the error signals that drive inference and learning in a cortical hierarchy?
Starting from maximum-likelihood estimation, \textcite{wilmes2025uncertaintymodulated} derive that the relevant error signal is not simply the difference between a sensory stimulus $s$ and its predicted mean $\mu$, but this difference weighted by the precision, i.e., the inverse variance, of the prediction.
Consequently, the same physical mismatch produces a smaller error signal when observations in the current context are expected to be highly variable, and a larger error signal when the variability is expected to be low.
\end{sloppypar}

The proposed cortical implementation combines subtractive and divisive inhibition (\cref{fig:uncertainty}a).
Because neuronal firing rates are non-negative and layer 2/3 pyramidal cells generally have low baseline firing rates \parencite{niell2008highly}, positive and negative prediction errors are represented by separate populations of error neurons, following earlier predictive-coding proposals \parencite{rao1999predictive,keller2018predictive}.

Each representation neuron is coupled to a pair of such circuits.
Each circuit contains two inhibitory populations of \gls{sst} and \gls{pv} interneurons, which, over time, learn to encode the first two moments of the stimulus from the current (sensory) context.
\gls{sst} cells replicate either the input $s$ or the prediction of its mean $r = \hat\mu$, while the \gls{pv} cells learn to predict the variance $\sigma^2$ of the stimulus from the context using the difference $s - \hat\mu$ as a teacher.
The pyramidal error neurons then compare the stimulus and its predicted mean through subtractive inhibition within a dedicated dendritic compartment~\parencite[see also][]{attinger2017visuomotor}.
The resulting dendritic mismatch signal is then divisively modulated by \gls{pv} inhibition close to the soma, thereby scaling the dendritic error by the expected variance of the prediction:
$e^+=\frac1{\sigma^2}(s-\hat\mu)^+$ and $e^-=\frac1{\sigma^2}(s-\hat\mu)^-$, for positive and negative errors respectively.
This \gls{upe} can now be used by basal synapses of the representation neuron to learn stimulus representations from contextual predictions, similarly to how predictions are learned in the models from \cref{sec:explicit}.
Over time, its prediction converges to the true mean $r \to \mu$, but with a context-dependent learning rate; thus, it has the ability to only update its predictions when it receives high-certainty sensory information.
To this end, dendritic and somatic compartments contribute distinct operations: dendrites support the comparison between predicted and observed input, whereas perisomatic gain control implements uncertainty weighting.

As an alternative to the microcircuit model proposed in \textcite{wilmes2025uncertaintymodulated}, \textcite{hertaeg2025uncertainty} show how the variance of sensory inputs can be estimated directly from the activity of the error neurons.
These estimates are not learned, but rather updated directly through neuronal dynamics, allowing hierarchical circuits to dynamically track both sensory and prediction uncertainty.

\paragraph{Biology}
The model proposes that \gls{sst} interneurons provide the subtractive component, whereas \gls{pv} interneurons provide the divisive component (\cref{fig:uncertainty}a).
In the positive prediction-error circuit, \gls{sst} activity represents the predicted mean and is subtracted from the sensory input.
In the negative prediction-error circuit, \gls{sst} activity instead represents the current stimulus, which is subtracted from an excitatory representation of the prediction.
\Gls{pv} activity, on the other hand, represents the expected variance in both circuits and divisively suppresses error-neuron responses.
Importantly, the mean and variance need not be supplied externally to the circuit. Local activity-dependent plasticity allows inputs conveying the current context to train \gls{sst} and \gls{pv} responses to reflect the mean and variance of the associated stimulus distribution, respectively. Because these statistics are stored in context-specific synaptic weights, the same inhibitory neurons can express different mean and variance estimates in different contexts. Under the assumptions of the model, \gls{pv} firing rates become proportional to the expected variance, providing the divisive signal required to generate \glspl{upe}.

When positive and negative \glspl{upe} drive the activity of representation neurons, uncertainty automatically controls the effective rate of learning. In a low-uncertainty context, prediction errors are amplified and the internal representation adapts rapidly. In a high-uncertainty context, the same deviations produce smaller updates, preventing the representation from tracking expected fluctuations in the sensory input. Extending the mechanism across cortical levels further allows sensory evidence and prior predictions to be combined according to their respective uncertainties. The model therefore links compartment-specific neuronal computations to context-dependent learning and Bayes-optimal inference, and predicts that identical sensory mismatches should evoke smaller layer 2/3 error responses in high-variance contexts, accompanied by cell-type-specific representations of the expected mean and variance.

\subsection{Dynamic uncertainty estimation}
Uncertainty varies dynamically with context, and good internal models of the world must account for this.
Normatively, this entails writing uncertainty not as a parameter but as a parameterized function of current neuronal activity.
Moreover, it has been argued in the predictive coding literature that uncertainty weighting of prediction errors is the backbone of attention \parencite{feldman2010attention}.
To be a satisfactory model of attention, this weighting 
must again vary dynamically with context on the time scale of neuronal dynamics, not synaptic learning.

When modeling uncertainty as selective modulation of dendritic compartments, the multiplicative gain associated to each dendritic compartment must thus be adjusted rapidly upon context switches.
Compartmentalization is essential here because the different inputs integrated into the soma need to be weighted independently (so-called segregation of inputs) and dynamically, at much faster time scales than typically affordable for synaptic learning.

As a possible mechanism for implementing such real-time uncertainty modulation, \textcite{granier2024confidence} consider dendritic inhibition and disinhibition by interneuron circuits.
These circuit motifs are ubiquitous in the cerebral cortex, the most widely known involving \gls{vip} and \gls{sst} interneurons \parencite{pi2013cortical}.
In the typical case of a two-point neuron representing the apical and somatic compartments of a deep cortical pyramidal cell, such a circuit notably controls the gain of apical-to-soma integration, as a function of top-down (contextual) inputs (\cref{fig:uncertainty}b).
Its dynamics can be derived from a normative model for predicting both the mean and the confidence (inverse variance) of sensory data.

A normative approach to learning synaptic weights in such networks now yields a twin set of error-correcting rules \parencite{granier2024confidence}.
Reminiscent of previous models, plasticity in weights $W$ between pyramidal representation neurons minimizes the confidence-weighted prediction error: $\Delta W \propto e\, r$, with $e = (s - \hat\mu) / \sigma^2$.
On the other hand, inhibitory plasticity in weights $A$ between interneurons and pyramidal neurons minimizes second-order errors $\delta$: $\Delta A \propto \delta \, r$.
These second-order errors are computed as a difference between confidence and performance, i.e., between the inverse uncertainty $\frac{1}{\sigma^2}$ and the prediction error $e^2$: $\delta= \frac{1}{\sigma^2} - e^2$.
This illuminates the diversity of potential error-correcting plasticity rules operating on dendritic information, and their capacity to learn not only first- but also higher-order moments of the input distribution.

Finally, and similarly to the \glspl{upe} discussed above, the effective learning rate of weights conveying classical (first-order) predictions again depends on the expected uncertainty.
Under low expected uncertainty (high confidence), prediction errors entail large changes, while under high expected uncertainty (low confidence), the same errors could be attributed to variability and entail smaller changes.

\paragraph{Biology}
Dynamic uncertainty estimation predicts that the signatures of uncertainty in neural circuits should change dynamically with higher-order representations.
If these models are correct,  cortico-cortical projections would compute either classical predictions or expected uncertainty, depending on the postsynaptic cell type.

Locally, signatures of expected uncertainty should be sought in inhibitory/disinhibitory interneuron circuits, ultimately leading to fast modulation.
For the computation of second-order errors in neuronal circuits, expected uncertainty signals must be compared to (subtracted from) actual error magnitudes, computed for example through the sum of positive and negative error signals onto basket cells as in~\textcite{hertaeg2025uncertainty}.
One hypothesis is that the somatic activity of a specific type of superficial pyramidal cells represents these second-order error signals, but \gls{vip} interneurons are also potential candidates \parencite{najafi2025unexpected}.

Finally, the normative framework derived in \textcite{granier2024confidence} suggests that errors afferent to a specific neuron should undergo two types of weighting.
If the neuron is not confident of its own prediction, i.e., if the expected variance of its prediction $\sigma^2$ is high, then its prediction error is not important, so it should be weighted divisively by $\sigma^2$.
At the same time, if afferent neurons from other areas are not confident about their own predictions of this neuron's activity, i.e., if their expected variances $\sigma_p^2$ are high, then the neuron should trust the received errors more than the received predictions, so it should weight the errors multiplicatively by $\sigma_p^2$.
In terms of circuits (and experimental predictions), this entails that the activity of neurons encoding errors and the apical dendrites of neurons encoding representations should be inversely modulated by top-down expected uncertainty.
This can also form a normative framework for explaining the gain of apical-to-soma integration in deep pyramidal cells, central to some modern theories of conscious processing \parencite{aru2020cellular}.
For a specific implementation, the laminar specificity of \gls{sst} activity \parencite{munoz2017layerspecific} and targets \parencite{naka2019complementary} offers a good candidate.

\section{Discussion}
\label{sec:discussion}

Taking stock, we have presented a selection of biologically inspired models of neuronal computation that use the flexibility offered by dendritic compartments to enable powerful and local plasticity rules.
The key pieces of information that are required by all of these rules are pre- and post- synaptic activities, as well as target or error information.
Synaptic weights are adjusted proportionally to their contribution to the misalignment between activities and targets, approximating gradient descent as a core underlying motif, and thereby inevitably yielding variants of the delta rule.
Using dendritic compartments, all of the factors in these learning rules find expression at the local level of each neuron.

From this taxonomy of models we can observe a relatively simple  principle at play, as manifested by the large overlap between models in terms of neuron morphology.
To wit, representation dendrites always learn to follow the activity at the soma, which is pulled away by some teacher \textendash{} either as a direct target or in the form of an error signal.
What individuates each model is mainly the network structure, which defines how representations and error (or teacher) signals are encoded and where they are transmitted.
In many cases this allows for specific experimental predictions, some of which have already been borne out by recent studies.

\paragraph{Plasticity and the real-world complexity of dendritic trees} Our perspective only requires the existence of few functionally distinct compartments.
The complexity of dendritic trees might suggest that this does exhaust their functional value.
While, as we argued early on, we should not fall victim to the naturalistic fallacy, it might well be that there is a good functional reason for this complexity.
Based on what these models tell us, we may speculate that an even more federated dendritic structure could serve the separate representation (and weighting) of the same principle of error correction, but with respect to different sensory modalities, context information, or, more generally, a useful separation in some abstract space that is not built-in by evolution, but that neurons can learn through experience and plasticity.
The well-documented mobility of synapses along the dendritic tree \parencite{bonhoeffer2002spine,bhatt2009dendritic} certainly points in this direction.

Moreover, plasticity mechanisms often rely on coincidence signals involving backpropagating action potentials or dendritic calcium spikes, which benefit from a degree of independence from somatic dynamics.
Consistent with this view, dendritic inhibition can locally gate these coincidence signals, effectively switching plasticity on or off in a pathway-specific manner \parencite{wilmes2016inhibition}.
Because the spatial location of inhibition determines which dendritic branches exhibit plasticity, different input streams can be learned independently without disrupting previously stored associations.

\paragraph{Spiking models}
Most models presented in this perspective are rate-based, whereas actual biological neurons communicate using discrete spikes (or bursts).
Indeed, even when the models are ostensibly spike-based, such as \gls{us} or \gls{burstprop}, they are usually implemented in terms of rates as tasks grow in complexity.
By their discrete nature, spiking models introduce additional difficulties regarding the encoding of error or target information.
Notably, several of the models subtract activities to determine errors, something that is trivial to implement when using a continuous encoding in terms of rates.
However, these subtractions become dangerous when dealing with spike signals, as small timing differences can turn an error signal from zero to having a large value, with potentially highly disruptive effects effects on learning.
This kind of issue is made worse in noisy systems such as biological networks or neuromorphic hardware \textendash{} the natural technological target of bio-inspired spiking models.

\paragraph{Modulatory mechanisms}

\begin{sloppypar}
Neurons have many ways to modulate their interactions beyond just spikes or firing rates, which in turn may strongly influence what information synapses can have local access to.
Such modulation can happen both pre-synaptically and post-synaptically.
Pre-synaptic modulation would include the already discussed spikes and bursts, but could potentially also involve spike shape effects \parencite{hoppa2014control,ramezani2018impacts}.
It would also include prospectivity with variable temporal horizons, which
can also adapt dynamically for more powerful temporal processing capabilities.
In parallel, temporal delays can hugely influence encoded information and these can be both pre- and post- synaptic: pre-synaptic delays due to axonal length and thickness, and post-synaptic delays because of synapse position on the (receiving) dendritic tree.
Finally, post-synaptic modulations also include all kinds of synaptic transfer functions, from short-term plasticity to stochastic transmission, plus further gating mechanisms mediated by neurotransmitters and neuromodulators.
These can all influence what information is available in which place of a compartmentalized neuron, and therefore can inform synaptic plasticity.
This zoo of multiplexing mechanisms is also relevant for information separation beyond compartments, as discussed further below.
\end{sloppypar}

\paragraph{Consequences for neuromorphic silicon}

In the vast majority of cases, neuromorphic networks capable of complex computation are not trained in situ, with local synaptic plasticity, but rather in a hardware-in-the-loop approach, with a separate computer carrying out the synaptic weight updates.
This holds for digital \parencite{esser2015backpropagation} and analog \parencite{schmitt2017neuromorphic,billaudelle2020versatile,goeltz2021fast,cramer2022surrogate} electronics, but also for other kinds of neuromorphic substrates 
\parencite{de2019machine}, with only rare exceptions \parencite{renner2024backpropagation} that are limited in scale.
Much of the advantage offered by physical implementations of neuronal networks is thus lost in the use of less efficient computing substrates and in the transport of information towards them.
We therefore argue in favor of a full physical instantiation of inference \emph{and} learning in neuromorphic silicon, and suggest that the neuron and synapse models discussed above can serve as blueprints towards achieving this goal.

While the specific argument for dendritic learning par\-a\-digms is indeed made possible by these recent advances, the idea of increasing efficiency by localizing information is not without precedent, and modern machine learning serves as a prime example.
Since the advent of deep learning, the increasing demand for compute has led to the development of tailored silicon~\parencite[e.g.,][]{jouppi2023tpu}.
Early on, it was recognized that in particular for data intensive applications such as the linear algebra involved in machine learning, increased locality of information can significantly increase efficiency~\parencite{kung1978systolic}.
Locality has since become a guiding principle in the structural design of efficient computing systems~\parencite{leiserson2020theres}.

Spatial and temporal locality becomes especially relevant for the real-time learning of patterns with temporal structure.
This is because any additional element of communication and storage induces a separation between the internal state of the system and its external stimuli, which cannot simply be frozen between consecutive computing steps in real-time settings.
At the algorithmic level, the dominant backpropagation-based paradigms for training neural networks are non-local in time, requiring the storage of trajectories during the forward pass for later use in gradient calculations in the backward pass. 
Neuromorphic implementations of real-time error transport and learning in networks of structured neurons, which leverage spatial locality to achieve temporal locality of learning, are therefore particularly interesting candidates for instantiating online learning in silico.
The resulting locality obviates the need for an external training algorithm, while at the same time saving memory and transport costs in the process.
Thus, as with other locality-preserving techniques, such models also hold substantial promise in curbing the exorbitant energy costs~\parencite{IEA2025EnergyAI} of modern machine learning systems.

It should also be noted that while the models presented here are designed to explain credit assignment in biological neuronal networks, engineered silicon does not come with the same constraints.
As has become apparent, implicitly or explicitly, in some of our above discussions, functionally useful features exist that biology cannot or at least does not realize; these include the likes of negative firing rates, multiple neuronal outputs, and the communication of multi-bit values, all of which are cheap and straightforward to implement in silico. 
The well-known adage about airplanes not having to flap their wings should also hold for neuromorphic architectures: given the flexibility to move beyond beyond biological constraints, such ideas should definitely be taken into consideration when they can clearly offer tangible benefits.

The neuromorphic community has certainly already taken note of the increased interest in neuronal compartmentalization and its functional consequences.
Examples of hardware platforms implementing multi-compartment circuits include not only simulated circuitry~\parencite{
    cartiglia2020errorpropagation,  %
    cartiglia2022stochastic,   %
    maryada2025canonical,  %
},
but also fully developed systems~\parencite{
    wang2011twodimensional,  %
    davies2018loihi,  %
    billaudelle2020versatile,  %
    pehle2022brainscales,  %
    richter2024dynapse2,  %
}.
However, current implementations are mostly limited to small scales, and demonstrations of true advantages of compartmentalized (as opposed to point) neurons remain scarce and specifically do not include on-chip training of elaborate tasks~\parencite{
    yang2020scalable,  %
    polykretis2020astrocytemodulated,  %
    kaiser2022emulating,  %
    park202322pjspike,  %
    uludag2024biorealistic,  %
    dietrich2025sequence,  %
    goeltz2025delgrad,
}.
This mismatch highlights that challenges such as noisy substrates~\parencite{semenova2019fundamental,billaudelle2022accurate} and spiking communication (see above) still need to be overcome, but is also a testament to the difficulty of embedding complex local plasticity into neuromorphic silicon in the first place.

\paragraph{Information separation: beyond compartments}

So far, we have argued for dendritic compartments as providing the physical substrate for storing separate pieces of information.
We conclude our discussion with a more critical perspective of this fundamental premise and discuss potential alternatives.

\begin{sloppypar}
Instead of encoding different quantities in different physical locations, they could be encoded by means of different physical carriers, such as ion concentrations.
Indeed, intracellular calcium concentration is a key factor driving synaptic plasticity \parencite{malenka1988postsynaptic,graupner2005stdp,graupner2012calciumbased}, where long-term potentiation is induced when the intracellular calcium concentration is above a given threshold, while long-term depression is induced below this threshold.
It would therefore not be unrealistic to see this difference between the calcium concentration and the threshold calcium concentration as an error signal. %
\end{sloppypar}

Yet another option would be to separate the information in different times or phases such as the wake and sleep phases in Boltzmann machines \parencite{ackley1985learning}.
During the wake phase, the output neurons are driven by sensory stimuli, while during the sleep phase, those neurons are only driven by internal representations.
The synaptic learning rule then simply computes the difference of pre-post correlations between those two phases, which represents the error between the sampled and the target distribution.
If the learning rate $\alpha$ is small enough, the weights can simply integrate those correlations in the different phases without the need to store them explicitly.

Finally, the separation of information could take place in the frequency domain.
Fast information carriers such as spikes, which can encode temporal information at the millisecond resolution, could be separated from slower information carriers such as membrane potentials, which fluctuate on time scales of tens of milliseconds and more.
From this perspective, a synapse could differentiate the (fast) back-propagating action potential from a slower \gls{psp}.
As an example, both the voltage triplet rule \parencite{clopath2010connectivity} and the original triplet rule \parencite{pfister2006triplets} explicitly build on this ingredient.
In both models, long-term potentiation depends on the product of a fast and a slow postsynaptic variable.
Later studies further take into account the propagation of the membrane potential along the dendritic tree \parencite{letzkus2006learning,graupner2012calciumbased}.

While it remains to be shown how any of these mechanisms can allow the kind of deep credit assignment and complex real-time modulation that the dendritic models are capable of, their presence in biology and their importance for plasticity is uncontroversial.
Whether as an alternative to the dendritic segregation of information, or as complementary, facilitating factors, a complete picture of deep credit assignment in the brain will need to take them into account.

\section*{Acknowledgements}

\begin{sloppypar}
We gratefully acknowledge funding from the European Union under grant agreement \#101147319 (EBRAINS 2.0; BvH, FB, JG, TG, PH, JJ, MAP).
KM thanks the Volkswagen Foundation for funding under the call ``NEXT--Neu\-ro\-mor\-phic Computing'', and the Swiss National Science Foundation (grant number 225643).
AG is supported by the Agence National de la Recherche (ANR-23-IACL-0008) and IN\-SERM.
KW is funded by the Swiss National Science Foundation Starting Grant (TMS\-GI3\_225\-811).
JG and MAP also thank the German Research Foundation for support through the Heidelberg STRUCTURES Excellence Cluster (Germany's Excellence Strategy EXC 2181/1-390\-900948).
JPP is supported by the Swiss National Science Foundation Grant ``Why spikes?'' (310030\_212247).
BvH, PH, SB and AG acknowledge the support of the Graduate School for Cellular and Biomedical Sciences (GCB) of the University of Bern.
JG and TG acknowledge the support of the Heidelberg Graduate School For Physics (HGSFP).
Last but not least, we owe a particular debt of gratitude for the ongoing support from the Manfred Stärk Foundation.
\end{sloppypar}

\renewcommand{\glossarypreamble}{\small}
\printglossaries{}

{
    \setcounter{biburllcpenalty}{7000}
    \setcounter{biburlucpenalty}{8000}
    \setcounter{biburlnumpenalty}{9000}
    \emergencystretch=2em
    \printbibliography
}

\end{document}